\documentstyle[pra,aps]{revtex}
\begin{document}

%\draft

%\preprint{DUKE-TH-94-76}

\title{Physics and Signatures of the Quark-Gluon
Plasma\footnote{Review article prepared for {\sl Report of Progress in
Physics}}}

\author{Berndt M\"uller}
\address{Department of Physics, Duke University, Durham NC 27708-0305}

\maketitle

\begin{abstract}
This is a critical review of the various observables that have been
proposed to signal the change from dense hadronic matter to a quark-gluon
plasma at high temperature or baryon density.  I discuss current models
of quark-gluon plasma formation in relativistic heavy ion collisions
and analyze the virtues and ambiguities of various signatures.
\end{abstract}
\bigskip\bigskip\bigskip

%\pacs{...}

%\newpage

\tableofcontents

\newpage
\section{Introduction}

The central goal of relativistic heavy ion physics is the discovery of
the new state of strongly interacting matter that is called the quark-gluon
plasma and the study of its properties \cite{r1}.  In order to accomplish
this goal, we need reliable signals for the formation of such a state.
This review tries to capture the essential ideas and the current status
of the theoretical studies on the most promising quark-gluon plasma signals.
The reader is encouraged to also consult the earlier reviews of Kajantie
and McLerran \cite{r2} and Singh \cite{r2a} and the Proceedings of the
Strasbourg Workshop \cite{r3}, which contain detailed accounts of various
signatures.

In order to shed light on the connections between the many proposed
quark-gluon plasma signatures I will here group them in five categories,
according to the physical properties of superdense hadronic matter to
which they are sensitive.  These are:  thermodynamic variables measuring
the equation of state; probes for chiral symmetry restoration; probes of
the color response function; probes of the electromagnetic response
function; and ``exotic'' signatures of the quark-gluon plasma.  Some
signals, e.g. strangeness production, may be sensitive to more than one
particular aspect.  In this case they have been grouped in the section
which I considered to be most relevant.

The review begins with a brief survey of our current picture of the
dynamical and structural properties of the quark gluon plasma, of its
creation and evolution.  It then proceeds to the analysis of the
various proposed signals and concludes with an assessment of their
merits and ambiguities.

\section{Brief Survey of Quark-Gluon Plasma Physics}

Experiments at Brookhaven and CERN have provided compelling evidence
that highly energetic collisions between heavy ions proceed through an
intermediate phase that is characterized by energy densities in excess
of 1 GeV/fm$^3$\cite{r4}.  This can be concluded from the analysis of
rapidity distributions of particle multiplicities and transverse
energy, as well as from the observed depletion of charmonium
production which requires the presence of a dense cloud of comovers.
It is less certain at this point whether this dense hadronic phase can
be considered as locally quasi-thermal, but theoretical studies indicate
that local thermalization of matter in relativistic heavy ion collision
actually proceeds more rapidly as the bombarding energy grows.
Estimates of the thermalization time, which are based on perturbative
QCD and the parton model \cite{r5,r6,r7}, as well as recent
nonperturbative studies of lattice gauge theory \cite{r8} yield values
below 0.5 fm/c at collider energies (above 100 GeV per nucleon in the
c.m. system).

The problem of the equation of state of superdense hadronic matter is
a fundamental issue of quantum chromodynamics.  Theory tells us that
it makes little sense to base a description of strongly interacting
matter at energy densities far above 1 GeV/fm$^3$ on a picture beginning
with noninteracting hadrons.  Rather it appears to be appropriate to
start from the fundamental constituents of hadrons, i.e. quarks and
gluons, because interactions between quarks and gluons become weak at
short distances.  This approach, anticipated in pre-QCD days by P.
Carruthers \cite{r9}, was pioneered by Collins and Perry \cite{r10},
Baym and Chin \cite{r11}, McLerran et al. \cite{r12}, Shuryak \cite{r13},
and others.  It is complicated by the fact that the QCD plasma retains
essentially nonperturbative aspects even at the highest densities because
of the existence of collective modes (plasmons) and the absence of
screening of long-range static color-magnetic forces in perturbation
theory.  Although plasma properties can be classified according to
orders of the QCD coupling constant $g$ (not $\alpha_s = g^2/4\pi$),
$g\ll 1$ does {\it not} hold even at the highest conceivable energies
($g\approx 0.5$ at the Planck scale).

\subsection{Structure of the Quark-Gluon Plasma}

The theory of the equation of state of the gluon plasma is
conceptually quite simple, because it is directly based on the
fundamental QCD Lagrangian
\begin{equation}
{\cal L}_{QCD} = {
\textstyle{1\over 4}} \sum_a F_{\mu\nu}^a
F^{a\mu\nu} + \sum_{f=1}^{N_f}
\overline{\psi}(i\gamma^{\mu}\partial_{\mu} - g\gamma^{\mu}\,
A_{\mu}^a \textstyle{{\lambda^a\over 2}} - m_f) \psi \label{1}
\end{equation}
where the subscript $f$ denotes the various quark flavors $u,d,s,c$,
etc., and the nonlinear glue field strength is given by

\begin{equation}
F_{\mu\nu}^a = \partial_{\mu} A_{\nu}^a - \partial_{\nu} A_{\mu}^a +
gf_{abc} A_{\mu}^b A_{\nu}^c. \label{2}
\end{equation}
QCD predicts a weakening of the quark-quark interaction at short
distances (or high momenta $Q^2$), because the one-loop series for the
gluon propagator yields a running coupling constant

\begin{equation}
g^2(Q^2) = {16\pi^2\over (11-{2\over 3}N_f) \ln (Q^2/\Lambda^2)}
\buildrel Q^2\to\infty \over \longrightarrow 0, \label{3}
\end{equation}
where $N_f$ is the number of active quark flavors.  The QCD scale
parameter $\Lambda$ is now quite well determined \cite{Lambda} to have
the value $\Lambda \approx$ 200 MeV.

Assuming that interactions of quark and gluons are sufficiently small
at high energy, the energy density $\varepsilon$ and pressure $P$ of a
quark-gluon plasma at temperature $T$ and quark chemical potential
$\mu_f$ can be calculated by thermal perturbation theory.  Neglecting
quark masses in first-order perturbation theory, the equation of state
is
\begin{eqnarray}
\varepsilon &=& \left( 1 - {15\over 4\pi} \alpha_s\right) {8\pi^2\over
15} T^4 + N_f \left( 1 - {50\over 21\pi} \alpha_s\right) {7\pi^2\over
10} T^4 + \nonumber \\
& & \sum_f \left( 1 - {2\over \pi} \alpha_s \right) {3\over
\pi^2} \mu_f^2 \left( \pi^2 T^2 + {1\over 2}\mu_f^2\right) + B,
\label{4}
\end{eqnarray}
where $B$ is the difference between the energy density of the
perturbative and the nonperturbative QCD vacuum (the bag constant).
The pressure is given by $P={1\over 3} (\varepsilon- 4B)$; the entropy
density is $s= (\partial P/\partial T)_\mu$.  One observes that
(\ref{4}) is essentially the equation of state of a gas of massless
particles with corrections due to the QCD trace anomaly and
perturbative interactions.  These are always negative, reducing the
energy density at given temperature by about a factor two when
$\alpha_s$ = 0.5.  The perturbative plasma phase becomes unstable at
temperatures below $T_c \approx 0.8 B^{1/4} \approx 170$ MeV
(for $\mu$ = 0), for standard values of the bag constant.

More reliable predictions concerning this phase transition can at
present only be obtained by numerical simulations of the QCD equation
of state on a finite discretized volume of space-time, usually referred
to as lattice gauge theory.  In this approach \cite{r33} one
approximately calculates the partition function for a discretized
version of the QCD Lagrangian (\ref{1}) by Monte-Carlo methods.  In
principle, this technique should accurately describe the quark-gluon
plasma as well as the hadronic phase but, in practice, its accuracy
especially at low temperature is severely limited by finite size
effects and other technical difficulties.  Where the numerical results
are most reliable, i.e. for the pure gluon theory without dynamical
quarks, the calculations predict a sudden jump in the energy density
at a certain temperature while the pressure rises more gradually.

When dynamical quarks are added, the picture becomes less clear for
two reasons.  One is that the calculations involving fermion fields on
the lattice are much more time consuming, and hence the numerical
results are less statistically meaningful and reliable.  Moreover, the
definition of quark confinement becomes rather fuzzy in the presence
of light quarks, because the color flux tube between two heavy quarks
can break by creation of a light quark pair:  $Q\overline{Q}\to
(Q\overline{q})(q\overline{Q})$.  E.g., highly excited states of
charmonium can break up into a pair of D-mesons.  Thus, in the
calculations the $q\overline{Q}$ potential does not rise linearly with
distance, but is effectively screened.

For massless dynamical quarks there exists a new order parameter, the
quark-antiquark condensate in the vacuum $\langle 0\vert\overline{q}
q\vert 0\rangle$.  When it assumes a nonzero value, chiral symmetry is
spontaneously broken, as can be seen as follows:  The scalar quark
density has the chiral decomposition $\overline{q}q = \overline{q}_L
q_R+\overline{q}_Rq_L$, hence the broken vacuum state contains pairs
of quarks of opposite chirality.  A left-handed quark, say, can
therefore annihilate on a left-handed antiquark in the vacuum
condensate, liberating its right-handed partner.  This process is
perceived as change of chirality of the free quark, which provides
exactly the same effect as a nonvanishing quark mass.  However, in
reality the mass of the light quarks $u,d$ is nonzero, and the chirality
of a light quark is never exactly conserved, even when the quark
condensate vanishes.

All one can do, therefore, is to look for sudden changes in the
distance at which color forces are screened, or in the quark
condensate.  If these are discontinuous, one deals with a phase
transition, otherwise with a possibly rapid, but continuous change of
internal structure as it occurs, e.g., in the transformation of an
atomic gas into an electromagnetic plasma.  The identification of the
nature of the phase change is complicated by finite size effects.  The
best published results, by the Columbia group \cite{r34,Chr92}, for a
$16^3\times 4$ lattice indicate a surprisingly strong dependence of
the phase diagram on the magnitude of the strange quark mass.  For the
physical mass $m_s\approx$ 150 MeV the presence of a discontinuity has
not been established, but a very rapid change in the energy density over
a small temperature range (about 10 MeV) is clearly seen in the
numerical results.

In the next order of the coupling constant $\alpha_s$, the gluon energy
density is found to be infrared divergent.  The physical reason for this
divergence is that gluon and quark degrees of freedom develop an effective
mass, which leads to screening of long-range color-electric forces.
Technically, the screening mass is obtained by summing an infinite chain of
one-loop insertions in the gluon propagator, which can be summed
analytically and yield a contribution to the gluon energy of order
$\alpha_s^{3/2}$ with a rather large coefficient \cite{r51}.

One obtains more insight into the properties of the interacting
quark-gluon plasma by considering the gluon propagator
$D_{\mu\nu}(k)$.  Because of gauge invariance, $k^{\mu}D_{\mu\nu}(k)
= 0$, it can be decomposed into longitudinal and transverse parts, which
are scalar functions of the variables $\omega = k^0$ and $k=\vert k\vert$.
These are most conveniently written in the form
\begin{equation}
D_L(\omega,k) =  {1\over \varepsilon_L(\omega,k)k^2}, \label{7a}
\end{equation}
\begin{equation}
D_T(\omega,k) = {1\over \varepsilon_T(\omega,k)\omega^2-k^2},
\label{7b}
\end{equation}
where the color-dielectric functions are given by \cite{r52}:
\begin{equation}
\varepsilon_L(\omega,k) = 1 + {g^2T^2\over k^2} \left[
1 - {\omega\over 1k} \ln \left( {\omega+k\over\omega-k}\right)\right];
\label{8a}
\end{equation}
\begin{equation}
\varepsilon_T(\omega,k) = 1-{g^2T^2\over 2k^2} \left[
1-\left( 1-{k^2\over\omega^2}\right) {\omega\over 2k} \ln \left(
{\omega+k\over\omega-k}\right)\right]. \label{8b}
\end{equation}
Several things are noteworthy about (\ref{7a}--\ref{8b}).
First they imply that static longitudinal color fields are screened:
\begin{equation}
D_L(0,k) = {1\over \varepsilon_L(0,k)k^2} = {1\over k^2+g^2T^2}. \label{9}
\end{equation}
The Debye length obviously is $\lambda_D = (gT)^{-1}$.  On the other
hand, (\ref{7a}--\ref{8b}) show that static transverse (magnetic)
color fields remain unscreened at this level of approximation.
The static magnetic screening length is of higher order in the
coupling constant; lattice gauge calculations \cite{r53} as well as
analytical studies \cite{r53a} have shown that $\lambda_M^{-1} =
Cg^2T$ with $C\approx$ 0.31 for SU(3) gauge theory.

For a finite frequency $\omega$ the in-medium propagators (\ref{7a},\ref{7b})
have poles corresponding to propagating collective modes of the glue
field.  The dispersion relation for the longitudinal mode:
\begin{equation}
\varepsilon_L(\omega,k) = 0, \label{10a}
\end{equation}
called the {\it plasmon}, has no counterpart outside the medium.  The
analogous relation for the transverse mode:
\begin{equation}
\varepsilon_T(\omega,k) = k^2/\omega^2 \label{10b}
\end{equation}
describes the effects of the medium on the free gluon.  The behavior
of both modes is remarkably similar.  For $k\to 0$ they yield an
effective plasmon mass
\begin{equation}
\omega_L,\omega_T \buildrel {k\to 0} \over {\longrightarrow}
m_g^* = {1\over\sqrt{3}} gT, \label{11}
\end{equation}
whereas for large momenta $(k\to\infty)$ one finds
\begin{equation}
\omega_L(k) \to k, \qquad \omega_T(k)\to \sqrt{k^2 + {1\over 2}
g^2T^2}. \label{12}
\end{equation}
For plasma conditions realistically attainable in nuclear collisions
$(T\approx$ 300 MeV, $\alpha_s \approx$ 0.3) the effective gluon mass
$m_g^*$ is of the order of the temperature itself.  We must conclude
therefore, that the notion of almost free gluons (and quarks) in the
high-temperature phase of QCD is quite far from the truth.

Let us discuss some consequences of these results:

1. The potential between two static color charges, such as two heavy
quarks, is screened in the quark-gluon plasma phase.  The Fourier
transform of eq. (\ref{9}) yields the potential
\begin{equation}
V_{Q\overline{Q}}(r) \sim {1\over r} e^{-r/\lambda_D} \label{13}
\end{equation}
with screening length $\lambda_D \approx$ 0.4 fm at $T\approx$ 250 MeV.
The screening of long-range color forces is, of course, the origin of
quark deconfinement in the high-temperature phase.  An important
consequence, to be discussed later, is the disappearance of the bound
states of a charmed quark pair $(c\bar c)$ in the quark-gluon plasma
\cite{r54}.

2.
The color screening at large distances cures most infrared divergences
in scattering processes between quarks and gluons.  A self-consistent
scheme implementing this mechanism has been devised by Braaten and
Pisarski \cite{r55}.  It involves the resummation of gluon loops
involving gluons with momenta of order $gT$ and has been shown to be
gauge invariant when vertex corrections are also taken into account.

3.
The finite effective gluon mass $m_g^*$ leads to the suppression of
long-wavelength gluon modes with $k \le gT$ in the quark-gluon
plasma.  As a result, the pressure is reduced.  We now have two
mechanisms that can be responsible for $P < {1\over 3}\varepsilon$:
an effective gluon mass $m_g^*$ and a nonvanishing vacuum energy $B$.
Fits \cite{r57,Pes94} to SU(3) lattice gauge theory results \cite{r56}
show that probably both mechanisms are at work.

Whereas the depletion of long-wavelength modes in the quark-gluon
plasma appears to be rather well established, its dynamical origin is
by no means clear.  If the above picture is correct, their population
is simply suppressed by the collective mass of the colored plasmons
On the other hand, it has also been speculated that low-momentum modes
in the QCD plasma are effectively reduced due to clustering into
color-singlet quasi-particles, which are remnants of hadronic states
above $T_c$.  Indeed, lattice gauge calculations have provided evidence
that color-singlet clusters can survive at high
temperature in spite of color screening \cite{r57a} (but see
\cite{r57b}).  A phenomenological model has been constructed which
ascribes the different behavior of $\varepsilon (T)$ and $P(T)$ near
to the postulated cluster structure of the quark-gluon plasma \cite{r57c}.
Several signals are expected to be very sensitive to the microscopic
structure of the QCD plasma at low momenta, e.g. production of
multistrange baryons or low-mass lepton pairs.  A better understanding
of the long-range structure of the quark-gluon plasma, and of the
interpretation of the pertinent lattice-QCD results, is therefore
of paramount importance.

\subsection{Properties of the Quark-Gluon Plasma}

\subsubsection{Thermalization}

All models of the formation of the quark-gluon plasma in nuclear
collisions require information about its rate of thermalization.  Does
it thermalize sufficiently fast, so that a thermodynamical description
makes sense?  In the picture based on quasi-free quarks and gluons
moving through the plasma, thermalization proceeds mainly via two-body
collisions, where the color force between the colliding particles is
screened.  The technically easiest way of looking at this is to consider
the quark (gluon) {\it damping rate} $\Gamma$, i.e. twice the imaginary
part of the quark (gluon) self energy.  For quarks and gluons of typical
thermal momenta $(p \approx T)$ one finds \cite{r55} with the help of the
techniques discussed in the previous section\footnote{(Note the additional
factor two compared to the standard definition, $\Gamma=2\gamma$).}
\begin{eqnarray}
\Gamma_Q &= &-{1\over 2p}\;\; \hbox{Im} \left[ \hbox{Tr}\Sigma(p)\right]
\approx {g^2\over 3\pi}T \left( 1+ \ln {1\over \alpha_s}\right),
\label{14a} \\
\Gamma_G &= &-2 \;\; \hbox{Im} [\omega_T(p)] \approx 0.54g^2T. \label{14b}
\end{eqnarray}
The larger factor for gluons simply reflects the fact that gluons have
a higher color charge than quarks and therefore scatter more often.

One can argue that a better way to look at thermalization is to
consider the rate of momentum transfer between particles, i.e.
weighting the differential scattering cross section by $\sin^2\theta$,
where $\theta$ is the scattering angle.  Here one finds \cite{r58}:
\begin{equation}
\Gamma_Q^{(\rm tr)} \simeq 2.3\alpha_s^2 T\; \ln {1\over \alpha_s};
\quad
\Gamma_G^{(\rm tr)} = 3\Gamma_Q^{(\rm tr)}. \label{15}
\end{equation}
The {\it transport rate} $\Gamma^{(\rm tr)}$ is closely related to the
shear viscosity of the quark gluon plasma.  Both approaches yield
quite similar numbers for values of the strong coupling constant in
the range $\alpha_s = 0.2-0.5$.  The characteristic equilibration
time, as defined as the inverse of the rates $\Gamma_i$ or
$\Gamma_i^{\rm (tr)}$ is:
\begin{equation}
\tau_G \approx 1\; \hbox{fm}/c, \quad \tau_Q \approx 3 \;\hbox{fm}/c,
\label{16}
\end{equation}
i.e. gluons thermalize about three times faster than quarks
\cite{r5}.  Initially, therefore, probably a rather pure glue plasma
is formed in heavy ion collisions, which then gradually evolves into a
chemically equilibrated quark-gluon plasma.  This picture is supported
by microscopic models of relativistic nuclear collisions, but rather
independent of the details entering into these models \cite{r6,r7}.
It is worthwhile pointing out one consequence:  during its hottest
phase the QCD plasma is thought to be mainly composed of gluons, which
are not accessible to electromagnetic probes, such as lepton pairs and
photons.  The gluon plasma can be best probed by strongly interacting
signals, such as charmed quarks or jets.

\subsubsection{Dynamical chaos}

In the early 1980's it was discovered by Matinyan, Savvidy and others
\cite{r66}  that classical Yang-Mills fields represent a chaotic dynamical
system in the extreme infrared limit.  The full space-time dynamics
can be investigated by evolving the gauge fields in real time on a lattice.
{}From a systematic study of the evolution of
randomly chosen field configurations Trayanov and M\"uller \cite{r67}
and Gong \cite{r68} concluded that nonabelian gauge fields are
characterized by a universal maximal Lyapunov exponent $\lambda_0$,
which for the SU(3) gauge theory obeys the equation $\lambda_0 \approx
{1\over 10} g^2 \langle E_p\rangle$, where $\langle E_p\rangle$
denotes the average energy per lattice plaquette.  For a thermalized
system $\langle E_p\rangle = {16\over 3}T$ holds in SU(3), hence the
Lyapunov exponent satisfies
\begin{equation}
\lambda_0 \approx 0.54 g^2T. \label{17}
\end{equation}
This corresponds to the rate with which, in the classical limit, a
small field perturbation grows.  It numerically coincides with the
thermal damping rate of long-wavelength plasmons (\ref{14b})
which describes the rate of decay of a deviation from thermal
equilibrium in perturbation theory.  The characteristic time $\tau_s =
\lambda_0^{-1} \approx \Gamma_G(0)^{-1}$ corresponds to the
thermalization of the most long-ranged modes.  For the temperature
range expected from a violent heavy ion collision this thermalization
time is less than 0.5 fm/$c$.

\subsection{Formation and Evolution of the Quark-Gluon Plasma}

\subsubsection{Formation of a Thermalized State}

Presumably, a quark-gluon plasma state can only be created in the
laboratory by collisions between two nuclei at very high energy.  But
how do the fully coherent parton wave functions of two nuclei in their
ground states evolve into locally quasi-thermal distributions of partons
as they are characteristic of the quark-gluon plasma state?  There are
mainly two approaches to this problem that have been extensively
investigated:  (a) QCD string breaking, and (b) the partonic cascade.

In the string picture developed from models of soft hadron-hadron
interactions, one assumes that nuclei pass through each other at collider
energies with only a small rapidity loss (on average about one unit),
drawing color flux tubes, or strings, between the ``wounded'' nucleons.
If the area density of strings is low (not much greater than 1 fm$^{-2}$)
they are supposed to fragment independently by quark pair production on
a proper time scale or order 1 fm/$c$.  Most realizations of this picture
are based on the Lund string model \cite{r69}, e.g. {\sc Fritiof}
\cite{r70}, {\sc Attila} \cite{r71}, {\sc Spacer} \cite{r72}, {\sc Venus}
\cite{r73}, QGSM \cite{r74} and RQMD \cite{r75}.  When the density of
strings grows further, at very high energy and for heavy nuclei, the
formation of ``color ropes''  instead of elementary flux tubes has been
postulated \cite{r76,r76a}.

Alternatively, a continuum description based on the Schwinger model of
(1+1)-dimensional QED with heuristic
back-reaction---``chromohydrodynamics''---has been invoked to describe
the formation of a locally equilibrated quark-gluon plasma
\cite{r77,Kag86,r78}.  One general aspect of these models is that initially
part of the kinetic energy of the colliding nuclei is stored in
coherent glue field configurations, which subsequently decay into
quark pairs.  The flux tubes carry no identifiable entropy.  The
entropy associated with a thermal state is produced in the course of
pair creation.  In particular, there is no distinction between gluon
and quark thermalization.

The parton cascade approach \cite{r78a}, which has precursors in the
work of by Boal \cite{r79}, Hwa and Kajantie \cite{r80}, Blaizot and
Mueller \cite{r81} and Eskola et al. \cite{r81a}, is founded on the
parton picture and renormalization-group improved perturbative QCD.
Whereas the string picture runs into conceptual difficulties at very
high energy when the string density becomes too large, the parton
cascade becomes invalid at lower energies, where most partonic
scatterings are too soft to be described by perturbative QCD.

We distinguish three-regimes in the evolution of an ultra-relativistic
heavy-ion collision.  Immediately after the Lorentz contracted nuclear
``pancakes'' have collided, scattered partons develop an incoherent
identity and evolve into a quasithermal phase space distribution by free
streaming separation of the longitudinal spectrum.  Rescattering of these
partons finally leads to the thermalization after a time probably much
less than 1 fm/$c$.  The thermalized quark-gluon plasma then evolves
according to the laws of relativistic hydrodynamics, until it has cooled
to the critical temperature $T_c \simeq 150-200$ MeV, where it begins to
hadronize.

The physics governing the evolution during the three states is
quite different.  When the two Lorentz-contracted nuclei collide, some
of their partons will scatter and then continue to evolve incoherently
from the remaining partons.  A parton-parton scattering can be
described by perturbative QCD if the momentum transfer involved is
sufficiently large.  It is not entirely clear where perturbative QCD
becomes invalid, but popular choices \cite{r85,r86,r87}
for the lower momentum cut-off are $p_T^{\rm min} \simeq 1.7-2$ GeV/c.
The time it takes for the scattered parton wave functions to
decohere from the initial parton cloud depends on the transverse
momenta of the scattered partons.  Usually, one argues that the
partons must have evolved at least one-half transverse wavelength
$\lambda_T = \pi/p_T$ away from their original position before they
can be considered as independent quanta.  The considerations
underlying this argument are similar to those for the
Landau-Pomeranchuk effect \cite{r60,r61,Gyu94}.

Complete calculations following the evolution of the parton
distributions microscopically until the attainment of thermal
equilibrium have been carried out by K. Geiger \cite{r6} who
found almost fully thermalized phase space distributions of
gluons in Au + Au collisions at RHIC energy $(E_{cm} =$ 100 GeV/u)
with $T\approx$ 325 MeV after proper time $\tau \approx$ 1.8 fm/c (see
Figure 10).  These results make it very likely that a quark-gluon
plasma will be produced in experiments at RHIC.

\subsubsection{Expansion and Chemical Equilibration}

Once the quark-gluon plasma has reached local thermal equilibrium, its
further evolution can be described in the framework of relativistic
hydrodynamics \cite{r83} .  The hydrodynamic equations for an
ultrarelativistic plasma with $P = {1\over 3}\varepsilon$ admit a
boost-invariant solution describing a longitudinally expanding
fireball with constant rapidity density.  When transverse expansion
effects are taken into account, longitudinal boost invariance is
partially destroyed, but the overall picture remains intact.  This
scenario has been extensively studied, and is described in several
publications and reviews \cite{r83a,r83b}, to which the interested
reader is referred.

Thermal equilibration does not always imply chemical equilibration at
the parton level.  A general result of all partonic cascade simulations
is that phase space equilibration occurs much faster for gluons than for
quarks.  As a result, the QCD plasma in its hottest stage is predominantly
a gluon plasma \cite{r5,r82}.  In the thermalized phase, this can be
described by different fugacities for gluons and quarks, $\lambda_G$ and
$\lambda_Q$, which evolve with time due to chemical reactions among partons
and the hydrodynamic longitudinal expansion of the quark-gluon plasma
\cite{r7,r84,Alam94}.

The chemical equilibration of heavier quarks, especially strange and
charmed quarks, has been investigated by many authors, because they
are thought to be excellent probes of the physical conditions in dense
hadronic matter.  Strange quarks are predicted to equilibrate on a
time scale of about 5 fm/$c$ \cite{r57,Ra82,Mat86,Ko86,Ba88}.
Even if the pure plasma phase does not live sufficiently long, almost
complete equilibration should occur during the mixed phase when the
quark-gluon plasma converts into a hadron gas \cite{Kaj86a}.
With the much higher initial energy densities predicted by the parton
cascade \cite{r6}, even charmed and bottom quarks may be abundantly
produced in the plasma phase \cite{Shor88,Mu82,Al93,Ge93a,Lin94}, but
it is quite unlikely that a complete phase space equilibration can be
reached for these heavy flavors.

\subsubsection{Hadronization}

An important aspect of the late evolution of the quark-gluon plasma is
its hadronization.  Mostly it is assumed that the plasma expands and
cools until it reaches the critical temperature $T_c\approx 170$ MeV
and then converts into a hadronic gas while maintaining thermal and
chemical equilibrium.  More detailed descriptions of the dynamics of
hadronization have been developed in connection with the problem of
strange hadron production \cite{Ko86,Ba88} and pion production
\cite{Mu85,Be88,Vi91}.

A totally different approach, explored more recently, consists in
following the partonic reactions at a microscopic level, until the
parton density has become sufficiently low to permit the formation of
individual hadrons \cite{Ge93b}.  A great deal is known about the
mechanism of final state hadron production in $e^+e^-$- and
NN-scattering but, unfortunately, we do not know whether this
knowledge applies to the hadronization of a quark-gluon plasma in bulk.

At present, it is hard to judge the merits of either approach on a
theoretical basis.  their validity and usefulness simply depends on
whether the microscopic processes during hadronization proceed
approximately at thermodynamical equilibrium or not.  There are
indications from strange hadron yields that hadronization may occur
quite suddenly \cite{Le93}.  The possibility of large deviations from
thermal equilibrium during the chiral phase transition has recently
attracted considerable interest, since it could lead to the formation
of ``disoriented'' chiral condensate states that would manifest
themselves in unusual pion charge ratios.  This phenomenon will be
discussed in more detail later (section III.B.2).

\section{Quark-Gluon Plasma Signatures}

Theoretical speculations about the nature of the quark-gluon plasma
would remain academic if there existed no experimental tools to
observe its formation and to study its properties.  There are several
arguments against the existence of unambiguous signatures of a
quark-gluon plasma formed in heavy-ion collisions.  The size of the
plasma volume is expected to be small, at most a few fermis in diameter,
and it does not live long, maybe between 5 and 10 fm/$c$.  Furthermore,
all signals emerging from the plasma are processed through, or receive
background from, the hot hadronic gas phase that follows the
hadronization of the plasma phase.  Nonetheless, a remarkable wealth of
ideas has been put forward in the past decade as to how the identification
and investigation of the short-lived quark-gluon plasma phase could be
accomplished.

It is impossible to present an exhaustive review of quark-gluon plasma
signatures here, and we must concentrate on the most promising ones.
Many omitted details can be found in other reviews \cite{r2,r2a} and
conference proceedings \cite{r3}.  Since signals are designed to be
sensitive to certain physical properties of the quark-gluon plasma
phase, they will be classified here in the following way:  (1) signals
sensitive to the equation of state; (2) signals of chiral symmetry
restoration; (3) probes of the color response function (including
deconfinement); (4) probes of the electromagnetic response function;
(5) various other signatures that escape simple classification.

\subsection{Probes of the Equation of State}

The basic idea behind this class of signatures is the identification
of modifications in the dependence of energy density $\epsilon$,
pressure $P$, and entropy density $s$ of superdense hadronic matter on
temperature $T$ and baryochemical potential $\mu_b$.  One wants to
search for a rapid rise in the effective number of degrees of freedom,
as expressed by the ratios $\epsilon/T^4$ or $s/T^4$, over a small
temperature range.  These quantities would exhibit a discontinuity in
the presence of a first-order phase transition, at least if we were
dealing with systems of infinite extent.  More realistically, we can
expect a steep, step-like rise as predicted by recent lattice
simulations \cite{r34}.

Of course, one requires measurable observables that are related to the
variables $T$, $s$, or $\epsilon$.  It is customary to identify those
with the average transverse momentum $\langle p_T\rangle$, and with
the rapidity distribution of the hadron multiplicity $dN/dy$, or
transverse energy $dE_t/dy$, respectively \cite{Ho82}.  One can then, in
principle, invert the $\epsilon$-$T$ diagram and plot $\langle
p_T\rangle$ as function of $dN/dy$ or $dE_t/dy$.  If there occurs a
rapid change in the effective number of degrees of freedom, one
expects an S-shaped curve, whose essential characteristic feature is
the saturation of $\langle p_T\rangle$ during the persistence of a
mixed phase, later giving way to a second rise when the structural
change from color-singlet to colored constituents has been completed.
Detailed numerical studies in the context of the hydrodynamical model
have shown that this characteristic feature is rather weak in realistic
models \cite{r83a,Ge87,Ka92}.  The strength of the transverse flow signal
can be enhanced by studying higher moments of the momentum distribution,
or heavier hadrons such as baryons \cite{Br86}.  It has been shown that
a pion gas is probably too dilute to accelerate baryons efficiently
\cite{Lev91,Prak92}, hence an observed collective baryon flow must be
attributed to an earlier phase of the evolution of the hadronic fireball.
The transverse flow signal could also be enhanced by the existence of a
sharp phase boundary or detonation wave \cite{Gy84,Ba85,Sei85,Cl86,Bi93}.

Flow effects in transverse momentum spectra are notoriously difficult
to detect.  While model calculations \cite{Am91} and phenomenological
analyses \cite{Schn92} of measured transverse momentum distributions
of particles \cite{Schu90} at presently accessible energies (up to 200
GeV/nucleon) point to the presence of transverse flow, the phenomenon
is far from established in this energy range.  Most likely, a credible
analysis will require the study of anisotropic flow patterns allowing
for the determination of the reaction plane in noncentral collisions.
The techniques \cite{Gy82,Da83} that have proved very successful at
lower energies may also be useful in the domain of ultrarelativistic
energies \cite{Ol92}.

In order to trace these effects in nuclear collisions one probably
has to vary the beam energy in small steps, as will be feasible at
RHIC.  In nucleon-antinucleon collisions, however, one may make use of
the existence of large fluctuations in the total multiplicity even for
central N-N collisions.  This idea has been probed by the E-735
collaboration at Fermilab \cite{Alex90}, who found a continued rise of
$\langle p_T\rangle$ for antiprotons and hyperons with multiplicity,
exceeding 1 GeV/$c$ for the most violent events $(dN/dy>20$).  An analysis
in terms of a schematic, scaling flow model \cite{Lev91} showed that
these experimental results would be consistent with transverse flow,
but this hydrodynamic model is at variance with the small source sizes
deduced from two-pion correlations \cite{Alex93}.  It is more likely
that minijets are the origin of the apparent transverse flow
\cite{Wa92}.  The importance of the minijet component is expected to
grow with increasing center-of-mass energy.  Minijets are predicted to
be a major mechanism of energy deposition in heavy ion reactions at
collider energies \cite{r81a,Es91,Wa91}, as discussed before.  It is
also worthwhile mentioning that an increase in the average transverse
momentum of produced particles with multiplicity is observed even in
$e^+e^-$ annihilations into hadrons \cite{Del92}, where it can be
explained by minijet branching processes \cite{Blo92}.

Models of the space-time dynamics of nuclear collisions need
independent confirmation, especially concerning the correctness of
their geometrical assumptions.  Such a check is provided by identical
particle interferometry, e.g. of $\pi\pi$, KK, or NN correlations,
which yield information on the reaction geometry.  By studying the
two-particle correlation function $F_2(q) =1+C_2(q)$ in different
directions of phase space, it is possible to obtain measurements of
the transverse and longitudinal size, of the lifetime, and of flow
patterns of the hadronic fireball at the moment when it breaks up
into separate hadrons \cite{Pr84,Ber89}.  The transverse sizes found
in heavy ion collisions \cite{Ba89,Ab92,Bog93,Ak93}, as well as in N-N
collisions at high multiplicity \cite{Alex93}, are larger than the
radii of the incident particles clearly exposing the fact that
produced hadrons rescatter before they are finally emitted.  Recent
theoretical work has pointed out the importance of the finite lifetime
of the fireball \cite{Cso94} and of shadowing effects \cite{Chu94}.
Since interferometric size determinations will be possible on an
event-by-event basis when Pb or Au beams become available, the
correlation of global parameters like $\langle p_T\rangle$ and $dN/dy$
with the fireball geometry may be performed on individual collision events.

Vector meson decays into lepton pairs also may provide information
about flow patterns and the lifetime of a mixed phase of quark-gluon
plasma droplet and hadronic matter.  These effects will be discussed
in section III.D.

\subsection{Signatures of Chiral Symmetry Restoration}

\subsubsection{Strangeness Enhancement}

The most often proposed signature for a restoration of spontaneously
broken chiral symmetry in dense baryon-rich hadronic matter are
enhancements in strangeness and antibaryon production.  the basic
argument in both cases is the lowering of the threshold for production
of strange hadrons and baryon-antibaryon pairs.  An optimal signal is
obtained by considering strange antibaryons, which combine both effects
\cite{Ra82,Ra82a}.

An enhancement of strange particle production in nuclear collisions
has been observed in many experiments
\cite{Od90,Ha91,Bi92,Aba93,An92,Ba93}.  However it has also been
learned that such an enhancement alone does not make a reliable
signature for the quark-gluon plasma.  Strange particles, especially K
mesons and $\Lambda$ hyperons can be copiously produced in hadronic
reactions before the nuclear fireball reaches equilibrium.  This mechanism
seems to work very efficiently at presently accessible energies
\cite{Ma89,Ni89}.  The processes leading to enhanced K and $\Lambda$
production have been studied in detail in the framework of hadron cascade
models, where it was found that most of the enhancement comes from
meson-baryon reactions initiated by mesons produced in first collisions,
which have a strongly nonthermal spectrum \cite{Ma89}.

However, such calculations based on the notion of a cascade of binary,
sequential hadron interactions fail to describe the full specter of data
obtained from heavy ion collisions at 200 GeV/nucleon.  The
enhancement of $\Lambda$-hyperon production over a wide rapidity range
\cite{Ba93}, and especially the observed strongly enhanced yields of
$\bar\Lambda, \bar\Xi$, and even $\Omega,\bar\Omega$ hyperons
\cite{Aba93,An92} require severe modifications of the hadronic cascades.
The data can be explained, if more exotic mechanisms such as color rope
formation \cite{r76a,So92}, multiple string breaking \cite{We93}, or decaying
multi-quark droplets \cite{Ai93a,Ai93} are incorporated into the cascade
models.  Of course, such mechanisms should be considered as phenomena
associated with the onset of quark-gluon plasma formation.

A simple thermal model based on the assumption of a rapidly decaying
quark-gluon plasma with strange quark fugacity $\lambda_s$ near unity
$(\mu_s=0)$ and light quark fugacity $\lambda_q \approx 1.5$
$(\mu_q/T\approx 0.4)$ describes the observed strange hadron multiplicities
rather well \cite{Raf92,Let94}.  However, a thermally, but not quite
chemically equilibrated hadron gas can also explain the hadron yield
ratios \cite{Dav91,Let92}.  The equilibrated fireball model, especially
for a quark-gluon plasma, provides a simple explanation for the relatively
large abundance of anti-hyperons, such as $\bar\Lambda,\bar\Xi$, and
$\bar\Omega$.  The value $\mu_s=0$ is natural in an environment where
quarks are deconfined and a baryon-antibaryon asymmetry does not affect
strange quarks, but it is rather unnatural in a hadronic environment at
nonvanishing net baryon density \cite{Raf92}.  The plasma model is
favored when the entropy/baryon content of the disintegrating
fireball is also considered \cite{Le93,So93}.  However, certain
inconsistencies remain.  It is not completely clear how a dense, hot
fireball with temperature around 200 MeV can explosively disintegrate
without final state interactions that modify the hadron yields.  Furthermore,
the fireball assumption contradicts the conclusions about the reaction
dynamics derived from more sophisticated hadronic cascade models that
predict a larger entropy per baryon \cite{So93a}.  On the other hand,
one can argue that the existing data point to a sequential freeze-out
scenario for strange and nonstrange hadrons \cite{Su94}.  The hadron
yields may also be sensitive to structural details of the quark-hadron
phase transition \cite{Bar93}.

It has been argued that strange particles, and especially antibaryons,
could also be produced more abundantly, if their masses are modified
in the hadronic phase due to medium effects \cite{Ko91}.  The mass of
K-mesons can be substantially lowered at finite baryon number density
(this could even lead to kaon condensation \cite{Kap86,Ne87,Po91}), and
the effective mass of antibaryons would be substantially reduced if
mean-field meson theories could be applied to the description of dense
hadronic matter.  It remains to be seen, though, whether models which
predict an enhancement in the $\bar\Lambda/\pi$ ratio by an order of
magnitude \cite{Scha91} can provide a consistent description of the many
other aspects of a heavy ion collision.

Strangeness enhancement may also affect the $\phi$-meson channel
\cite{Shor85}.  This was indeed observed in experiments at 200
GeV/nucleon ($S+U$ and $O+U)$ collisions) where a clear enhancement in
the $\phi/(\rho+\omega)$ yield ratio compared with $p+U$ interactions was
found \cite{Ba91}.  This enhancement can be explained on the basis of
quark-gluon plasma formation as well as by purely hadronic scenarios
\cite{Ko90,Gr91}.

In summary, the analysis of the implications of strange hadron abundances
is complicated by the fact that strangeness carrying hadrons interact
strongly.  Information carried by strange hadrons about their original
source may be lost in final state interactions.  Although these
interactions have been modeled in considerable detail
\cite{Ko86,Ba88}, many predictions are quite model dependent.  The
lack of experimental information about many interactions involving
strange hadrons is also an impeding factor \cite{Ai93}.  Furthermore,
it has been recognized \cite{Bar93} that the dynamical and structural
aspects of the quark-hadron transition must be much better understood
before conclusions about the existence of a quark-gluon plasma can be
safely drawn from hadron yields.  Despite these obstacles, however, strange
particle yields provide some of the most stringent tests for dynamic
models of ultrarelativistic nuclar collisions \cite{Koch91}.

\subsubsection{Disoriented Chiral Condensates}

A direct and virtually unambiguous signal for the restoration of
chiral symmetry in nuclear collisions could come from domains of
disoriented chiral condensate.  These correspond to isospin singlet,
coherent excitations of the pion field, and would decay into neutral
and charged pions with the probability distribution
\begin{equation}
P\left( {N_{\pi^0}\over N_{\pi}}\right) \sim {1\over 2}
\sqrt{{N_{\pi}\over N_{\pi_0}}}. \label{18}
\end{equation}
Although the average ratio $\langle N_{\pi^0}/N_{\pi}\rangle = {1\over
3}$ as required by isospin symmetry, final states with a large surplus
of charged pions over neutral pions, as they were observed in Centauro
events \cite{Centauro}, would occur with significant probability
\cite{Lam84,Pr93}.

The most likely origin of a coherent low-energy excitation of the pion
field would be a collective isosinglet excitation of the Goldstone boson
field $(\sigma,\vec\pi)$ associated with the spontaneous breaking of
chiral symmetry, which can be described as a nonlinear wave in the sigma
model \cite{Ans89,Bl92}.  Such a wave can be excited by the growth of local
instabilities during the transition from the chirally restored
high-temperature phase of QCD to the low-temperature phase, in which
chiral symmetry is broken \cite{Bj92,Kow92}.  These instabilities are a
direct consequence of the instability of the chirally symmetric ground state
below the temperature $T_c$ of the chiral phase transition.  The growth of
long wavelength modes in the chiral order parameter then occurs quite
naturally, if the transition proceeds out of equilibrium \cite{Ra93}.

Numerical calculations have shown that the coherence length of these
collective excitations remains quite small, of order 1 fm, if the chiral
field gets too far out of equilibrium \cite{Ga93a}.  As the coherence
length is inversely proportional to the growth rate of the instabilities
\cite{Bo93,Be93}, larger domains of coherently excited pion field may
emerge if the chiral order parameter is somewhat, but not too much, away
from its equilibrium \cite{Ga93b}.  Of course, such a scenario is a priori
more likely to occur in a relativistic heavy ion collision, because the
chiral transition is accompanied by a large change in the energy density.
The need to absorb the latent heat reduces the speed with which the
temperature falls in the vicinity of $T_c$, implying that the system
cannot get too far out of equilibrium until the chiral transition is
completed.  Boost invariant models of the dynamics of the chiral order
parameter have been studied \cite{Kl93,Ko93,Hu93,Bl94} with the
conclusion that coherent domains of the order of $3-5$ fm can be formed
in the cooling process \cite{Klu94,Asa94}.

It is quite unlikely that a Bose condensed state could be formed
simply by isentropic expansion of a dense pion gas, as this
contradicts simple entropy considerations \cite{Gr93}.  It has also been
shown \cite{Co94} that even quite large deviations from isospin
neutrality would not significantly destroy the signature provided by
the distribution (\ref{18}).  The observation of pion charge ratios
$N_{\pi^0}/N_{\pi}$ significantly different from ${1\over 3}$, or
nonzero charge correlations \cite{Gr93}, would therefore be a very
strong and direct signature of the chiral phase transition.

\subsection{Color Response Function}

The basic aim in the detection of a color deconfinement phase
transition is to measure changes in the color response function
\begin{equation}
\Pi_{\mu\nu}^{ab} (q^2) = \int d^4xd^4y\; e^{iq(x-y)}\langle
j_{\mu}^a(x) j_{\nu}^b(y)\rangle, \label{19}
\end{equation}
where $j_{\mu}^a(x)$ is the color current density.  Although this
correlator is not gauge invariant (except in the limit
$q\to 0$), its structure can be probed in two ways: (1) the screening
length $\lambda_D^2 \delta^{ab} = \Pi_{00}^{ab}(0)$ leads to
dissociation of bound states of a heavy quark pair $(Q\bar Q)$, and
(2) the energy loss $dE/dx$ of a quark jet in a dense medium is
sensitive to an average of $\Pi_{\mu\nu}^{ab}(q^2)$ over a wide range
of $q$.

\subsubsection{Quarkonium Suppression}

The suppression of $J/\psi$ production \cite{r54} is based on the
insight that a bound state of a $(c\bar c)$ pair cannot exist when the
color screening length $\lambda_D\approx 1/gT$ is less than the bound state
radius $\langle r_{J/\psi}^2\rangle^{1/2}$.  Thus a $c\bar c$ pair
formed by fusion of two gluons from the colliding nuclei cannot bind
inside the quark-gluon plasma \cite{Mehr88}.  Lattice simulations of
SU(3) gauge theory \cite{De86,Ka86} show that this condition should be
satisfied slightly above the deconfinement temperature $(T/T_c>1.2)$.
The screening length appears to be even shorter when dynamical fermions
are included in the lattice simulations \cite{Ka88}.  In addition,
the $D$-meson is expected to dissociate in the deconfined phase, lowering
the energy threshold $\Delta E^*$ for thermal break-up of the $J/\psi$
into two $D$-mesons.  The combination of these two effects leads to a
rapid rise of the dissociation probability beyond $T_c$, reaching unity
at about 1.2 $T_c$ \cite{Bl91}.  Excited states of the $(c\bar c)$
system, such as $\psi'$ and $\chi_c$, are even more rapidly dissociated
and should disappear as soon as the temperature exceeds $T_c$.  For
the heavier $(b\bar b)$ system similar considerations apply, although
shorter screening lengths are required than for the charmonium states
\cite{KaSa91}.  The dissociation temperature of the $\Upsilon$ ground
state is predicted to be around 2.5 $T_c$, that of the larger $\Upsilon'$
state around 1.1 $T_c$.

Owing to its finite size, the formation of a $(c\bar c)$ bound state
requires a time of the order of 1 fm/$c$.  This formation
time can be understood classically, as the time required for the
$(c\bar c)$ pair to get separated by a distance equal to the average
diameter of the bound state, or quantum mechanically, as the time needed
for the various eigenstates of the system to decohere
\cite{Cer90,Huf90,Thews90}.  The $J/\psi$ may still survive, if it escapes
from the region of high density and temperature before the $c\bar c$ pair
has been spatially separated by more than the size of the bound state
\cite{r54}.  This will happen either if the quark-gluon plasma cools very
fast, or if the $J/\psi$ has sufficiently high transverse momentum $p_T\ge$
3 GeV/c.  The phenomenology of the $J/\psi$ suppression due to quark-gluon
plasma formation has been extensively modeled, and its dependence on total
transverse energy and nuclear size, as well as on transverse momentum,
has been studied by many authors
\cite{Bl91,Bl87a,Ka88a,Ruu88,Chu88,Sat88,Mat88,Fta89,Rop88,Raha89,Lie91,Gaz91}.
The details of $J/\psi$ suppression near $T_c$ are quite complicated and
could require a rather long lifetime of the quark-gluon plasma state before
becoming
clearly visible because there are precursor effects \cite{Hi90}.

On the other hand, the $J/\psi$ may also be destroyed in a hadronic
scenario by sufficiently energetic collisions with comoving hadrons
\cite{Ga88,Vo88}, leading to dissociation into a pair of $D$-mesons.
This mechanism has been carefully analyzed \cite{Ga88b,Ga90,Vo91}.
Moreover, $J/\psi$ production is also ``suppressed'' in hadron-nucleus
collisions, i.e. the cross section for $hA\to J/\psi$ grows only as
$A^{\alpha}$ with $\alpha\approx 0.93 < 1$ \cite{Alde91}, as well as
in $\mu A$ collisions \cite{Ama91}.  Comprehensive analyses of the
available data have led to the conclusion that several effects
contribute to the observed suppression at small $x$:  nuclear
shadowing of gluons, initial state scattering of partons resulting in
a widened transverse momentum distribution, and final state absorption
on nucleons \cite{Vo91,Gup92,Vo92,Don93}.

These mechanisms do not explain, however, the full extent of $J/\psi$
suppression observed in nucleus-nucleus collisions by the NA38
experiment at CERN \cite{NA38}.  Additional absorption  of $J/\psi$ on
hadronic comovers (pions and resonances) with a density around
1/fm$^3$ and an absorption cross section $\sigma_{\rm abs}\approx 3$
mb can provide a good global description of the NA38 data
\cite{Gavin93}.  An equally good description of the present data can
be obtained by invoking the plasma dissociation mechanism
\cite{Shen93}.  The presence of absorption by comovers is also
supported by the observed stronger suppression of the $\psi'$
state \cite{NA38}, but this observation could be readily explained
by hadronic comovers \cite{Gavin93}.  Similar considerations, comparing
$J/\psi$ suppression with the observed $\phi$-meson enhancement \cite{Ba91}
also lead to the conclusion that hadronic comovers and a quark-gluon
plasma can describe the available data equally well \cite{Ko90}.  In view
of this persistent ambiguity, the observed $J/\psi$ suppression cannot at
present be considered as evidence for the formation of a quark-gluon plasma
in nuclear collisions.  Whether this ambiguity can be releived at higher
(collider) energies where $\Upsilon$, $\Upsilon'$ suppression can also be
studied, as has been argued in \cite{KaSa91}, remains to be seen.  In
spite of these ambiguities, heavy vector meson suppression is a clear
indicator of the presence of a high-density environment in the central
rapidity region formed in relativistic heavy ion collisions.

\subsubsection{Energy Loss of a Fast Parton}

Another possible way of probing the color structure of QCD matter is by the
energy loss of a fast parton (quark or gluon).  The mechanisms are
similar to those responsible for the electromagetic energy loss of a
fast charged particle in matter, i.e. energy may be lost either by
excitation of the penetrated medium or by radiation.

The energy loss of a fast quark in the quark-gluon plasma was first studied in
perturbative QCD \cite{Bj82} and later in the framework of transport
theory \cite{Sve88}.  The connection between energy loss of a quark
and the color-dielectric polarizability of the medium can be established
in analogy to the theory of electromagnetic energy loss
\cite{Tho91,Mro91,Koi91}.  The magnitude of the energy loss is
proportional to the strong coupling constant $\alpha_s^2$.
Different authors obtain a stopping power between 0.4 and 1 GeV/fm for
a fast quark \cite{Tho91,Mro91,Koi91}, which is somewhat smaller than
the energy loss of a fast quark in nuclear matter.

Although radiation is a very efficient energy loss mechanism for
relativistic particles, it is strongly suppresed in a dense medium by the
Landau-Pomeranchuk effect \cite{r60}.  In the case of QCD this effect
has recently been analyzed comprehensively \cite{r61}.  The suppression of
soft radiation \cite{Gyu94} limits the radiative energy loss to about
1 GeV/fm \cite{Gyu91}.  Adding the two contributions it thus appears that
the stopping power of a fully established quark-gluon plasma is probably
higher than that of hadronic matter.

\subsection{Electromagnetic Response Function}

Photons and lepton pairs \cite{Shu78,MT85,Raha91} are in many respects
the cleanest signals for the quark-gluon plasma because they probe the
earliest and hottest phase of the evolution of the fireball, and are not
affected by final state interactions.  Their drawbacks are the rather small
yields and the relatively large backgrounds from hadronic decay
processes, especially electromagnetic hadron decays.  Electromagnetic
signals probe the structure of the electromagnetic current response
function
\begin{equation}
W_{\mu\nu} (q^2) = \int d^4x\;d^4y\; e^{iq(x-y)} \langle j_{\mu}(x)
j_{\nu}(y)\rangle. \label{20}
\end{equation}
In the hadronic phase, $W_{\mu\nu}(q^2)$ is dominated by the $\rho^0$
resonance at 770 MeV, whereas perturbative QCD predicts a broad
continuous spectrum above twice the thermal quark mass
$m_q=gT/\sqrt{6}$ in the high temperature phase.  At $q^2 < $ 100 MeV
collective modes are predicted to exist in both phases.  The collective
plasma excitation, the ``plasmino'' \cite{Wel91}, has a higher effective
mass than the collective $\pi^+\pi^-$ mode \cite{Gale91}; but both lie in
a region where their signal will probably be overwhelmed by background, in
particular by lepton pairs emitted from non-collective modes, and by
the Dalitz pair background \cite{Braa90}.  On the other hand, lepton
pair production in the quark-gluon plasma may be sensitive to
preequilibrium phenomena, e.g. collective plasma oscillations of large
amplitude \cite{Asa91}.  Such oscillations are known to occur in the
framework of the chromo-hydrodynamic model \cite{r77,Kag86,r78} where the
collision energy is first stored in a coherent color field which later
breaks up into $q\bar q$ pairs.  The ensuing collective flow effects could
strongly enhance the production of lepton pairs of high invariant mass.

\subsubsection{Lepton Pairs}

Lepton pairs have been considered as probes of the quark-gluon plasma
since the earliest days
\cite{Fe76,Shu78,Dom81,Kaj81,Chin82,Dom83,McL85,Cley86,Hwa85}.  Many of these
original calculations concentrated on lepton pairs emitted with
invariant masses in the energy range below the $\rho$-meson mass.
However, with an improved understanding of the collision dynamics and
the hadronic backgrounds \cite{Cley91}, it has since become clear that
lepton pairs from the quark-gluon plasma can probably only be observed
for invariant masses above $1-1.5$ GeV \cite{Kaj86,Kaj87,Ruu91,Ruu92}.
Assuming thermalization of the quark-gluon plasma on a time scale of
about 1 fm/$c$, the thermal dilepton spectrum is superseded by Drell-Yan
pairs from first nucleon-nucleon collisions at invariant masses around
$2-2.5$ GeV.

The great progress in understanding the mechanisms of thermalization made
recently has completely changed this outlook.  It has become clear that
the yield of high-mass dileptons critically depends on, and provides a
measure of, the thermalization time \cite{Kap92}.  Lepton pairs from the
equilibrating quark-gluon plasma may dominate over the Drell-Yan background
up to masses in the range $5-10$ GeV, as predicted by the parton cascade
\cite{Gei93} and other models of the early equilibration phase of the
nuclear collision \cite{Shu93,Kam92,Kaw92}.  If lepton pairs can be
measured above the Drell-Yan background up to several GeV of invariant
mass, the early thermal evolution of the quark-gluon phase can be
traced in a rather model independent way \cite{Stri94}.  Dileptons from
charm decay are predicted to yield a substantial contribution to the total
dilepton spectrum and could, because of their different kinematics, provide
a useful determination of the total charm yield \cite{Vogt93}.  This
could serve as an indirect probe of the partonic preequilibrium phase,
where the total charm yield is enhanced due to rescattering of gluonic
partons \cite{Mu82,Lin94}, if the direct background is sufficiently well
understood \cite{Sar94}.

Although they do not constitute direct probes of the quark-gluon
plasma, lepton pairs from hadronic sources in the invariant mass range
between 0.5 and 1 GeV could be valuable signals of the dense hadronic
matter formed in nuclear collisions.  The suggestion \cite{Sie85} that
the disappearance of the $\rho$-meson peak in the lepton pair mass
spectrum would signal the deconfinement transition has recently been
revived \cite{Sei92}.  The basic idea is to utilize the fact that the
quark-gluon plasma phase should exhibit the higher temperature than the
hadronic phase, and therefore lepton pairs from the quark-gluon plasma
should dominate at high $p_T$ over those originating from hadronic
processes.  Note that this reasoning may become inconclusive when one
allows for collective transverse flow.  Because of its larger mass, the
$\rho$-meson spectrum is much more sensitive to the presence of flow than
the quark spectrum in the quark-gluon plasma phase.  Kataja et al.
\cite{Kat92} have shown that the $\rho$-meson peak becomes more prominent
at large $p_T$, even if a quark-gluon plasma phase exists temporarily.

Nonetheless, the lepton pairs from $\rho$-meson decay can be a very
useful tool for probing the hadronic phase of the fireball.  Heinz and
Lee \cite{Hei91} have pointed out that the $\rho$-peak is expected to
grow strongly relative to the $\omega$ and $\phi$ peaks in the
electon pair mass spectrum, if the fireball lives substantially longer
than 2 fm/$c$.  This occurs because of the short average lifetime of the
$\rho$ (1.3 fm/$c$), so that several generations of thermal $\rho^0$
mesons would contribute to the spectrum.  In the limit of a very
long-lived fireball the ratio of lepton pairs from $\rho^0$ and
$\omega$ decays would approach the ratio of their leptonic decay
widths.  The $\rho/\omega$ ratio can therefore serve as a fast
``clock'' for the fireball lifetime.

The widths and positions of the $\rho,\omega$, and $\phi$ peaks should
also be sensitive to medium-induced changes of the hadronic mass
spectrum, especially to precursor phenomena associated with chiral
symmetry restoration \cite{Barz91,Chan92,Ko92,Herr92,Hat93}.  The
general conclusion, however, is that modifications of the peak
positions are probably small except in the immediate vicinity of the
phase transition.  Changes are predicted to occur sooner, if the hadronic
phase contains an appreciable net baryon density.  On the other hand,
the increase in the width of the $\phi$-meson due to collision
broadening is substantial \cite{Hag93}.  This could serve as a measure
of the density of the mixed phase.  A change in the K-meson mass also
would affect the width of the $\phi$ meson \cite{Lis91,Bi91}.  A double
$\phi$ peak in the lepton pair spectrum would be indicative of a
long-lived mixed phase \cite{Asa93}.

\subsubsection{Direct Photons}

Direct photons, the second electromagnetic probe of dense matter,
compete with the formidable background from the decays of pseudoscalar
mesons.  Reconstruction and subtraction of these decays with sufficient
accuracy is difficult, but remarkable progress in this respect has
been made \cite{WA80}.  At collider energies, the background from
decays is predicted to extend to large transverse momenta due to the
influence of minijets \cite{Sri92}, but this contribution may be
strongly suppressed by minijet rescattering and thermalization.

Even if decay photons can be subtracted, there remains the competition
between radiation from the quark-gluon plasma and from the hadronic
phase.  In contrast to the lepton-pair spectrum the hadronic radiation
spectrum is not concentrated in a single narrow resonance.  The dominant
source of photons from the thermal hadron gas is the
$\pi\rho\to\gamma\rho$ reaction \cite{KLS}, in which the very broad
$a_1$-meson may be an important contribution \cite{Xiong92}.  In the quark
phase the Compton scattering process $gq\to\gamma q$ dominates.  Infrared
singularities occuring in perturbation theory are cured by taking into
account medium effects \cite{KLS,Bai92}.  The result is that a hadron
gas and a quark-gluon plasma in the vicinity of the critical
temperature $T_c$ emit photon spectra of roughly equal intensity and
similar spectral shape \cite{KLS}.  However, a clear signal of photons
from the quark-gluon plasma would be visible for transverse momenta
$p_T$ in the range $2-5$ GeV/$c$ if a very hot plasma is initially formed
\cite{Stri94,Sri92,Cha92}.  The photon spectrum in the $p_T$ range
$1-2$ GeV/$c$ is mostly emitted from the mixed phase \cite{Sei93}.
Transverse flow effects make the separation of the contributions from the
different phases more difficult \cite{Alam93}, and destroy the
correlation between the slope of the photon spectrum in the intermediate
$p_T$ range and the temperature of the mixed phase \cite{Neu94}.

Photon pairs are of interest, because they are emitted more abundantly
than lepton pairs of the same invariant mass, and because they probe
components of the electromagnetic response function (\ref{20}) with
different spin \cite{Yo87,Red87}.  Moreover, the photon pair
correlations in momentum space can---in principle---be utilized for a
density interferometric analysis that retains information about the
quark-gluon plasma phase \cite{Mak89,Sri93a}.  The photon pair
correlations are very sensitive to transverse flow effects and could be
utilized to analyze the dynamics of the hadronization transition.

\subsection{Metastable Quark Matter}

There exist general arguments for the possible metastability of cold
quark matter with a high content of strange quarks
\cite{Bod71,FM78,Chin79,Wit84}, which are supported by calculations in
the framework of the MIT-bag model
\cite{Far84,Kas93,Gil93,Mad93,Des93}.  These models also predict that
strange quark matter could remain metastable at finite temperature
\cite{Lee93,Cha93a}.  Whereas there exist strong astrophysical limits
on the presence of absolutely stable quark matter \cite{Mad88}, the
existence of metastable drops of strange quark matter (strangelets) is
not excluded experimentally.  Estimate of the lifetime of such objects
against weak decay yield values of the order of $10^{-7}-10^{-6}$ s
\cite{Hei92}.

These predictions are of interest for relativistic heavy ion collisions,
because a high density of strange quarks is produced \cite{Craw94}.
There exists a mechanism for the separation of strange quarks and
antiquarks due to the evaporation of K-mesons from a baryon-rich
quark-gluon plasma \cite{Liu84,Grei87,Shaw92}.  Since this strangeness
disillation mechanism requires a nonvanishing baryon number density,
it should be especially effective a presently accessible energies
where baryons are scattered into the central rapidity region.  So far,
experimental searches have been negative
\cite{Bar90,Aoki92,Bor94,Long94,App94}, but highly improved
experiments are in progress.  The strangeness distillation mechanism
can be probed by density interferometry with neutral kaons
\cite{Grei89,Gyu92a}.  The observation of multiple strange hypernuclei
\cite{Scha93}, though interesting in itself, would provide strong
evidence against the metastability of small drops of strange quark
matter.

\section{Summary}

Experimental detection of a quark-gluon plasma in relativistic heavy
ion reactions at collider energies requires a combination of signals,
which probe different aspects of the high-temperature phase of QCD.
It is first necessary to establish the main reaction mechanism, to
obtain experimental information on the initial conditions, and to
determine the lifetime of the hot, dense fireball.  Electromagnetic
probes, charm yield, and density interferometry are promising tools
for this purpose.  It would be especially interesting to demonstrate
the preponderance of gluonic degrees of freedom in the initial entropy
production, as predicted by parton cascade models.  If the thermalization
times are as short, and the initial densities as high as currently
predicted, the observation of a state consisting of quasifree quarks and
gluons should be possible with little ambiguity.

The existence of a rather long-lived mixed phase at the quark-hadron
phase transition should be visible in flow effects and in the lepton-pair
spectrum.  Strange baryons are an excellent probe of dense baryon-rich
matter.  The observed enhancements are perhaps the best indication, so
far, that hadronic reaction models are insufficient.  The observation of
disoriented chiral domains would clearly demonstrate the presence of the
chiral phase transition.  Metastable strangelets, if produced and found,
would provide unassailable evidence for the existence of quark matter.

The primary goal of relativistic heavy ion physics remains the exploration
of the reaction mechanism at high energies.  Is there a transition from
the low energy regime, where reactions can be successfully described as
interacting hadronic cascades, to a high-energy regime, where quark and
gluon constituents provide a much simpler description of the first few
fm/$c$ of the reaction?  Do the pomeron-dominated soft strong interactions
disappear at sufficiently high energy, giving way to bulk interactions
that can be much more economically described in the framework of
perturbative QCD?  An experimental demonstration of these features would
constitute, in itself, the discovery of the quark-gluon plasma.  The fact
that hadronic cascade models are beginning to fail and to become
inconsistent at the highest currently accessible energies \cite{Wer94}
provides reason for optimism that this transition may be near.

\section*{Acknowledgments}

This work was supported in part by grant DE-FG05-90ER40592 from the
U.S. Department of Energy.


\begin{references}
\bibitem{r1} For an introduction to this field, see e.g.:
B. M\"uller, {\sl The Physics of the Quark-Gluon Plasma, Lecture Notes
in Physics}, Vol. 225 (Springer-Verlag, Berlin-Heidelberg 1985);
L. McLerran, {\sl Rev. Mod. Phys. {\bf 58}}, 1021 (1986); {\sl
Quark-Gluon Plasma}, edited by R. C. Hwa (World Scientific, Singapore,
1991).

\bibitem{r2} K. Kajantie and L. McLerran, {\sl Ann. Rev. Nucl. Part.
Sci. {\bf 37}}, 293 (1987).

\bibitem{r2a} C. P. Singh, {\sl Phys. Rep. {\bf 236}}, 147 (1993).

\bibitem{r3} {\sl Quark-Gluon Plasma Signatures}, edited by V. Bernard et al.
(Editions Frontieres, Paris, 1990).

\bibitem{r4} J. Stachel and G. R. Young, {\sl Ann. Rev. Nucl. Part.
Sci. {\bf 42}}, 537 (1992).

\bibitem{r5} E. Shuryak, {\sl Phys. Rev. Lett. {\bf 68}}, 3270 (1992).

\bibitem{r6} K. Geiger, {\sl Phys. Rev. {\bf D46}}, 4965 and 4986
(1992); {\sl Phys. Rep.} (in print).

\bibitem{r7} T. S. Bir\'o, E. van Doorn, B. M\"uller, M. H. Thoma, and
X. N. Wang, {\sl Phys. Rev. {\bf C48}}, 1275 (1993).

\bibitem{r8} T. S. Bir\'o, C. Gong, B. M\"uller, and A. Trayanov, {\sl
Int. J. Mod. Phys. {\bf C5}}, 113 (1994).

\bibitem{r9} P. Carruthers, {\sl Collective Phenomena {\bf 1}}, 147
(1973).

\bibitem{r10} J. C. Collins and M. Perry, {\sl Phys. Lett. {\bf 34}},
1353 (1975).

\bibitem{r11} G. Baym and S. A. Chin, {\sl Phys. Lett. {\bf 62B}},
241 (1986);  S. A. Chin, {\sl Phys. Lett. {\bf 78B}}, 552 (1978).

\bibitem{r12} B. Friedman and L. McLerran, {\sl Phys. Rev. {\bf D17}},
1109 (1978);  L. D. McLerran, {\sl Phys. Rev. {\bf D24}}, 450 (1981).

\bibitem{r13} E. V. Shuryak, {\sl Phys. Rep. {\bf 61}}, 71 (1980).

\bibitem{Lambda} See: {\sl Review of Particle Properties 1994}, section 25,
{\sl Phys. Rev. {\bf D50}}, 1173 (1994).

\bibitem{r33} M. Creutz, {\sl Quarks, Gluons and Lattices} (Cambridge
University Press, Cambridge, 1983).

\bibitem{r34} F. R. Brown, et al., {\sl Phys. Rev. Lett. {\bf 65}},
2491 (1990).

\bibitem{Chr92} N. H. Christ, {\sl Nucl. Phys. {\bf A544}}, 81c (1992).

\bibitem{r51} J. I. Kapusta, {\sl Nucl. Phys. {\bf B148}}, 461 (1979);
T. Toimela, {\sl Z. Phys. {\bf C17}}, 365 (1983).

\bibitem{r52} V. P. Silin, {\sl Sov. Phys. JETP {\bf 11}} 1136 (1960);
V. V. Klimov {\sl Sov. Phys. JETP {\bf 55}}, 199 (1982); H. A. Weldon,
{\sl Phys. Rev. {\bf D26}}, 1394 (1982).

\bibitem{r53} A. Billoire, G. Lazarides, and Q. Shafi, {\sl Phys.
Lett. {\bf 103B}}, 450 (1981); T. A. DeGrand and D. Toussaint, {\sl
Phys. Rev. {\bf D25}}, 526 (1982); G. Lazarides and S. Sarantakos,
{\sl Phys. Rev. {\bf D31}}, 389 (1985).

\bibitem{r53a} T. S. Bir\'o and B. M\"uller, {\sl Nucl. Phys. A {\bf
561}}, 477 (1993).

\bibitem{r54} T. Matsui and H. Satz, {\sl Phys. Lett. {\bf 178B}}, 416
(1986).

\bibitem{r55} R. D. Pisarski, {\sl Phys. Rev. Lett. {\bf 63}}, 1129
(1989); E. Braaten and R. D. Pisarski, {\sl Phys. Rev. {\bf D42}},
2156 (1990).

\bibitem{r57} T. Bir\'o, P. L\'evai, and B. M\"uller, {\sl Phys. Rev.
{\bf D42}}, 3078 (1990).

\bibitem{Pes94} A. Peshier, B. K\"ampfer, O. P. Pavlenko, and G. Soff,
Univ. Dresden preprint (1994); D. H. Rischke, preprint CU-TP-649,
Columbia University (1994).

\bibitem{r56} A. Ukawa, {\sl Nucl. Phys. {\bf A498}}, 227c (1989).

\bibitem{r57a} C. DeTar and J. Kogut, {\sl Pys. Rev. Lett. {\bf 59}},
399 (1987); C. Bernard et al., {\sl Phys. Rev. Lett. {\bf 68}}, 2125
(1992);  S. Schramm and M. C. Chu, {\sl Phys. Rev. {\bf D48}}, 2279 (1993).

\bibitem{r57b} C. Bernard et al. (MILC collaboration), preprint
UUHEP-92-10, $\langle$hep-lat/9211036$\rangle$.

\bibitem{r57c} D. Rischke, J. Schaffner, M. Gorenstein, A. Sch\"afer,
H. St\"ocker, and W. Greiner, {\sl Z Phys. {\bf C56}}, 325 (1992);
{\sl Phys. Lett. {\bf B278}}, 19 (1992).

\bibitem{r58} M. H. Thoma, {\sl Phys. Lett. {\bf B269}}, 144 (1991);
G. Baym, H. Monien, C. J. Pethick, and D. G. Ravenhall, {\sl Phys.
Rev. Lett. {\bf 64}}, 1867 (1990).


\bibitem{r66} S. G. Matinyan, G. K. Savvidy, and N. G.
Ter-Arutyunyan-Savvidy, {\sl Sov. Phys. JETP {\bf 53}}, 421 (1981);
{\sl JETP Lett. {\bf 34}}, 590 (1981); B. V. Chirikov and D. L.
Shepelyanskii, {\sl JETP Lett. {\bf 34}}, 163 (1981); {\sl Sov. J.
Nucl. Phys. {\bf 36}}, 908 (1982); G. K. Savvidy, {\sl Nucl. Phys.
{\bf B246}}, 302 (1984).  See also: T. S. Bir\'o, S. G. Matinyan, and
B. M\"uller, {\sl Chaos and Gauge Theories} (World Scientific,
Singapore, in print) and references therein.

\bibitem{r67} B. M\"uller and A. Trayanov, {\sl Phys. Rev. Lett. {\bf
68}}, 3387 (1992).

\bibitem{r68} C. Gong, {\sl Phys. Lett. {\bf 298}}, 257 (1993).

\bibitem{r69} B. Anderson, G. Gustafson, G. Ingelman, and T.
Sj\"ostrand, {\sl Phys. Rep. {\bf 97}}, 31 (1983).

\bibitem{r70} B. Nilsson-Almquist and E. Stenlund, {\sl Comp. Phys.
Comm. {\bf 43}}, 387 (1987).

\bibitem{r71} M. Gyulassy, preprint CERN-TH-4784 (1987, unpublished).

\bibitem{r72} T. Cs\"org\"o, J. Zim\'anyi, J. Bondorf, and H.
Heiselberg, {\sl Phys. Lett. {\bf B222}}, 115 (1989).

\bibitem{r73} K. Werner, {\sl Z. Phys. {\bf C42}}, 85 (1989).

\bibitem{r74} N. S. Amelin, K. K. Gudima, and V. D. Toneev, {\sl Yad.
Fiz. {\bf 51}}, 512 (1990).

\bibitem{r75} M. Sorge, H. St\"ocker, and W. Greiner, {\sl Nucl. Phys.
{\bf A498}}, 567c (1989);  {\sl Ann. Phys. {\bf 192}}, 266 (1989).

\bibitem{r76} T. S. Bir\'o, H. B. Nielsen, and J. Knoll, {\sl Nucl.
Phys. {\bf B245}}, 449 (1984).

\bibitem{r76a} H. Sorge, M. Berenguer, H. St\"ocker, and W. Greiner,
{\sl Phys. Lett. {\bf B289}}, 6 (1992).

\bibitem{r77} A. Bialas and W. Czy\'z, {\sl Phys. Rev. {\bf D31}}, 198
(1985); {\sl Nucl. Phys. {\bf B267}}, 242 (1986).

\bibitem{Kag86} S. Kagiyama, A. Nakamura, and A. Minaka, {\sl Prog.
Theor. Phys. {\bf 75}}, 319 (1986).

\bibitem{r78} K. Kajantie and T. Matsui, {\sl Phys. Lett. {\bf B164}},
373 (1985); G. Gatoff, A. K. Kerman, and T. Matsui, {\sl Phys. Rev.
{\bf D36}}, 114 (1986); M. Asakawa and T. Matsui, {\sl Phys. Rev. {\bf
D43}}, 2871 (1991); G. Gatoff, preprint ORNL/CCIP/91/24, Oak Ridge,
1991.

\bibitem{r78a} K. Geiger and B. M\"uller, {\sl Nucl. Phys. {\bf
B369}}, 600 (1992).

\bibitem{r79} D. Boal, {\sl Phys. Rev. {\bf C33}}, 2206 (1986).

\bibitem{r80} R. C. Hwa and K. Kajantie, {\sl Phys. Rev. Lett. {\bf
56}}, 696 (1986).

\bibitem{r81} J. P. Blaizot and A. H. Mueller, {\sl Nucl. Phys. {\bf
B289}}, 847 (1987).

\bibitem{r81a} K. J. Eskola, K. Kajantie, and J. Lindfors, {\sl Phys.
Lett. {\bf B214}}, 613 (1988).

\bibitem{r60} A. B. Migdal, {\sl Sov. Phys. JETP {\bf 5}}, 527 (1957).

\bibitem{r61} A. H. S\o renson, {\sl Z. Phys. {\bf C53}}, 595 (1992).

\bibitem{Gyu94} M. Gyulassy and X. N. Wang, {\sl Nucl. Phys. {\bf B420}},
583 (1994).

\bibitem{r83} J. D. Bjorken, {\sl Phys. Rev. {\bf D27}}, 140 (1983).

\bibitem{r83a} H. von Gersdorff, L. McLerran, M. Kataja, and P. V.
Ruuskanen, {\sl Phys. Rev. {\bf D34}}, 794 (1986); M. Kataja, P. V.
Ruuskanen, L. McLerran, and H. von Gersdorff, {\sl Phys. Rev. {\bf
34}}, 3755 (1986).

\bibitem{r83b} J. P. Blaizot and J. Y. Ollitraut, in {\sl Quark-Gluon
Plasma}, ed. R. C. Hwa (World Scientific, Singapore, 1991) p. 393.

\bibitem{r82} K. Geiger and J. I. Kapusta, {\sl Phys. Rev. {\bf D47}},
4905 (1993).

\bibitem{r84} L. Xiong and E. V. Shuryak, {\sl Phys. Rev. {\bf C49}},
2207 (1994).

\bibitem{Alam94} J. Alam, S. Raha, and B. Sinha, {\sl Phys. Rev. Lett.
{\bf 73}}, 1895 (1994).

\bibitem{r85} T. Sj\"ostrand and M. van Zijl, {\sl Phys. Rev. {\bf
D36}}, 2019 (1987).

\bibitem{r86} X. N. Wang and M. Gyulassy, {\sl Phys. Rev. {\bf D44}},
3501 (1991).

\bibitem{r87} N. Abou-El-Naga, K. Geiger, and B. M\"uller, {\sl J.
Phys. {\bf G18}}, 797 (1992).

\bibitem{Ra82} J. Rafelski and B. M\"uller, {\sl Phys. Rev. Lett. {\bf
48}}, 1066 (1982); {\bf 56}, 2334(E) (1986).

\bibitem{Mat86} T. Matsui, B. Svetitsky, and L. McLerran, {\sl Phys.
Rev. {\bf D34}}, 783 and 2047 (1986).

\bibitem{Ko86} P. Koch, B. M\"uller, and J. Rafelski, {\sl Phys. Rep.
{\bf142}}, 167 (1986).

\bibitem{Ba88} H. W.  Barz, G. L. Friman, J. Knoll, and H. Schulz,
{\sl Nucl. Phys. {\bf A484}}, 661 (1988); {\sl Nucl. Phys. {\bf
A519}}, 831 (1990); {\sl Phys. Lett. {\bf B254}}, 315 (1991).

\bibitem{Kaj86a} K. Kajantie, M. Kataja, P. V. Ruuskanen, {\sl Phys.
Lett. {\bf B179}}, 153 (1986).

\bibitem{Shor88} A. Shor, {\sl Phys. Lett. {\bf B215}}, 375 (1988).

\bibitem{Mu82} B. M\"uller and X. N. Wang, {\sl Phys. Rev. Lett. {\bf
68}}, 2437 (1992).

\bibitem{Al93} T. Altherr and D. Seibert, {\sl Phys. Lett. {\bf
B313}}, 149 (1993); preprint CERN-TH-7038-93 (1993),
$\langle$nucl-th/9311028$\rangle$.

\bibitem{Ge93a} K. Geiger, {\sl Phys. Rev. {\bf D48}}, 4129 (1993).

\bibitem{Lin94} Z. Lin and M. Gyulassy, preprint CU-TP-638, Columbia
University (1994).

\bibitem{Mu85} B. M\"uller and J. M. Eisenberg, {\sl Nucl. Phys. {\bf
A435}}, 791 (1985).

\bibitem{Be88} G. Bertsch, M. Gong, L. McLerran, V. Ruuskanen, and E.
Sarkkinen, {\sl Phys. Rev. {\bf D37}}, 1202 (1988).

\bibitem{Vi91} A. P. Vischer, P. J. Siemens, and A. J. Sierk, {\sl Z.
Phys. {\bf A340}}, 315 (1991).

\bibitem{Ge93b} K. Geiger, {\sl Phys. Rev. {\bf D47}}, 133 (1993); K.
Geiger, preprint CERN-TH-7440/94, Geneva (1994),
$\langle$hep-ph/9409308$\rangle$.

\bibitem{Le93} J. Letessier, A. Tounsi, U. Heinz, J. Sollfrank, and J.
Rafelski, {\sl Phys. Rev. Lett. {\bf 70}}, 3530 (1993).


\bibitem{Ho82} L. van Hove, {\sl Phys. Lett. {\bf 118B}}, 138 (1982);
{\sl Z. Phys. {\bf C21}}, 93 (1983).

\bibitem{Ge87} H. von Gersdorff, {\sl Nucl. Phys. {\bf A461}}, 251c
(1987).

\bibitem{Ka92} M. Kataja, P. V. Ruuskanen, J. Letessier, and A.
Tounsi, {\sl Z. Phys. {\bf C55}}, 153 (1992).

\bibitem{Br86} J. P. Blaizot and J. Y. Ollitraut, {\sl Nucl. Phys.
{\bf A458}}, 745 (1986).

\bibitem{Lev91} P. L\'evai and B. M\"uller, {\sl Phys. Rev. Lett. {\bf
67}}, 1519 (1991).

\bibitem{Prak92} M. Prakash, M. Prakash, R. Venugopalan, and G. M.
Welke, {\sl Phys. Rev. Lett. {\bf 70}}, 1228 (1992).

\bibitem{Gy84} M. Gyulassy, K. Kajantie, H. Kurki-Suonio, and L. McLerran,
{\sl Nucl. Phys. {\bf B237}}, 477 (1984).

\bibitem{Ba85} H. W. Barz, L. P. Csernai, B. K\"ampfer, and B.
Luk\'acs, {\sl Phys. Rev. {\bf D32}}, 115 (1985).

\bibitem{Sei85} D. Seibert, {\sl Phys. Rev. {\bf D32}}, 2812; {\sl Phys.
Rev. {\bf D35}}, 2013 (1987).

\bibitem{Cl86} J. Cleymans, E. Nyk\"anen, and E. Suhonen, {\sl Phys.
Rev. {\bf D33}}, 2585 (1986); {\sl Z. Phys. {\bf C37}}, 51 (1987).

\bibitem{Bi93} N. Bil\'ic, J. Cleymans, E. Suhonen, and D. W. van
Oertzen, {\sl Phys. Lett. {\bf B311}}, 266 (1993).

\bibitem{Am91} N. S. Amelin, E. F. Staubo, L. P. Csernai, V. D.
Toneev, K. K. Gudima, and D. Strottman, {\sl Phys. Rev. Lett. {\bf
67}}, 1523 (1991).

\bibitem{Schn92} E. Schnedermann and U. Heinz, {\sl Phys. Rev. Lett.
{\bf 69}}, 2908 (1992).

\bibitem{Schu90} J. Schukraft, in ref. [4], p. 127.

\bibitem{Gy82} M. Gyulassy, K. A. Frankel, and H. St\"ocker, {\sl
Phys. Lett. {\bf 110B}}, 185 (1982).

\bibitem{Da83} P. Danielewicz and M. Gyulassy, {\sl Phys. Lett. {\bf
129B}}, 283 (1983).

\bibitem{Ol92} J. Y. Ollitraut, {\sl Phys. Rev. {\bf D46}}, 229 (1992);
{\sl Phys. Rev. {\bf D48}}, 1132 (1993).

\bibitem{Alex90} T. Alexopoulos et al. [E-735 collaboration], {\sl
Phys. Rev. Lett. {\bf 60}}, 1622 (1988); {\sl Phys. Rev. Lett. {\bf
64}}, 991 (1990); {\sl Phys. Rev. {\bf D48}}, 984 (1993); {\sl Phys.
Rev. {\bf D46}}, 2773 (1992).

\bibitem{Alex93} T. Alexopoulos et al. [E-735 collaboration], {\sl
Phys. Rev. {\bf D48}}, 1931 (1993).

\bibitem{Wa92} X. N. Wang and M. Gyulassy, {\sl Phys. Lett. {\bf
B282}}, 466 (1992).

\bibitem{Es91} K. J. Eskola, {\sl Nucl. Phys. {\bf A525}}, 393 (1991).

\bibitem{Wa91} W. N. Wang, in: {\sl Intersections between Particle and
Nuclear Physics}, AIP Conf. proceedings, vol. 243, p. 844.

\bibitem{Del92} DELPHI collaboration, {\sl Phys. Lett. {\bf B276}},
254 (1992).

\bibitem{Blo92} J. Blocki, M. Szczekowski, and G. Wilk, {\sl Acta
Phys. Polon. {\bf B23}}, 851 (1992).

\bibitem{Pr84} S. Pratt, {\sl Phys. Rev. Lett. {\bf 53}}, 1219 (1984);
{\sl Phys. Rev. {\bf D33}}, 1314 (1986).

\bibitem{Ber89} G. F. Bertsch, {\sl Nucl. Phys. {\bf A498}}, 173c
(1989).

\bibitem{Cso94} T. Cs\"org\"o, preprint LUNFDG-NFFL-7081, Lund
University (1994), $\langle$hep-ph/9409327$\rangle$.

\bibitem{Chu94} M. C. Chu, S. Gardner, T. Matsui, and R. Seki,
preprint MAP-175, Pasadena (1994), $\langle$nucl-th/9408005$\rangle$.

\bibitem{Ba89} A. Bamberger et al. [NA35 collaboration], {\sl Phys.
Lett. {\bf B203}}, 320 (1989); M. Lahanas et al [NA35 collaboration],
{\sl Nucl. Phys. {\bf A525}}, 327c (1991); D. Ferenc et al. [NA35
collaboration], {\sl Nucl. Phys. {\bf A544}}, 531c (1992).

\bibitem{Ab92} Abbott et al. [E-802 collaboration], {\sl Phys. Rev.
Lett. {\bf 69}}, 1030 (1992).

\bibitem{Bog93} H. Boggild et al. [NA44 collaboration], {\sl Phys. Lett.
{\bf B302}}, 510 (1993).

\bibitem{Ak93} Y. Akiba et al. [E-803 collaboration], {\sl Phys. Rev.
Lett. {\bf 70}}, 1057 (1993).

\bibitem{Ra82a} J. Rafelski, {\sl Phys. Rep. {\bf 88}}, 331 (1982);
{\sl Phys. Lett. {\bf B262}}, 333 (1991).

\bibitem{Od90} G. Odyniec, in ref. [4], p. 183.

\bibitem{Ha91} O. Hansen, {\sl Comm. Nucl. Part. Phys. {\bf 20}}, 1
(1992).

\bibitem{Bi92} H. Bialkowska, M. Ga\'zdzicki, W. Retyk, and E.
Skrzypczak, {\sl Z. Phys. {\bf C55}}, 491 (1992).

\bibitem{Aba93} S. Abatzis et al. [WA85 collaboration], {\sl Phys.
Lett. {\bf B316}}, 615 (1993); {\sl Nucl. Phys. {\bf A566}}, 225c
(1994); J. B. Kinson et al. [WA85 collaboration], {\sl Nucl. Phys.
{\bf A544}}, 321c (1992).

\bibitem{An92} E. Andersen et al. [NA36 collaboration], {\sl Phys. Lett
{\bf B294}}, 127 (1992); {\sl Phys. Rev. {\bf C46}}, 727 (1992); {\sl
Phys. Lett. {\bf B327}}, 433 (1994).

\bibitem{Ba93} M. Ga\'zdzicki et al. [NA35 collaboration], {\sl Nucl.
Phys. {\bf A566}}, 503c (1994); T. Alber et al. [NA35 collaboration],
preprint IKF-HENPG/1-94, Frankfurt (1994).

\bibitem{Ma89} R. Mattiello, H. Sorge, H. St\"ocker, and W. Greiner, {\sl
Phys. Rev. Lett. {\bf 63}}, 1459 (1989).

\bibitem{Ni89} N. N. Nikolaev, {\sl Z. Phys. {\bf C44}}, 645 (1989).

\bibitem{So92} H. Sorge, M. Berenguer, H. St\"ocker, and W. Greiner,
{\sl Phys. Lett. {\bf B289}}, 6 (1992).

\bibitem{We93} K. Werner and J. Aichelin, {\sl Phys. Lett. {\bf
B308}}, 372 (1993).

\bibitem{Ai93a} J. Aichelin and K. Werner, {\sl Phys. Lett. {\bf
B300}}, 158 (1993).

\bibitem{Ai93} J. Aichelin, in: {\sl Heavy Ion Physics at the AGS},
publication MITLNS-2158, MIT (1993), p. 169.

\bibitem{Dav91} N. J. Davison, H. G. Miller, R. M. Quick, and J.
Cleymans, {\sl Phys. Lett. {\bf B255}}, 105 (1991); D. W. von Oertzen,
N. J. Davidson, R. A. Ritchie, and H. G. Miller, {\sl Phys. Lett. {\bf
B274}}, 128 (1992).

\bibitem{Raf92} J. Rafelski, H. Rafelski, and M. Danos, {\sl Phys.
Lett. {\bf B294}}, 131 (1992); J. Rafelski, {\sl Nucl. Phys. {\bf
A544}}, 279 (1992).

\bibitem{Let94} J. Letessier, J. Rafelski, and A. Tounsi, {\sl Phys.
Lett. {\bf B321}}, 394 (1994).

\bibitem{Let92} J. Letessier, A. Tounsi, and J. Rafelski, {\sl Phys.
Lett. {\bf B292}}, 417 (1992).

\bibitem{So93} J. Sollfrank, M. Ga\'zdzicki, U. Heinz, and J. Rafelski,
preprint TPR-93-14, Univ. Regensburg (1993); J. Letessier, A. Tounsi,
U. Heinz, J. Sollfrank, and J. Rafelski, preprint PAR-LPTHE-92-27R,
Paris (1993).

\bibitem{So93a} H. Sorge, preprint LA-UR-93-1103, Los Alamos (1993),
unpublished.

\bibitem{Su94} E. Suhonen, J. Cleymans, K. Redlich, and H. Satz,
preprint UCT-TP-201-93, Univ. of Cape Town (1993),
$\langle$hep-ph/9310345$\rangle$.

\bibitem{Bar93} S. P. Baranov and L. V. Fil'kov, {\sl Z. Phys. {\bf
C57}}, 149 (1993).

\bibitem{Ko91} C. M. Ko, Z. G. Wu, L. H. Xia, and G. E. Brown, {\sl
Phys. Rev. Lett. {\bf 66}}, 2577 (1991).

\bibitem{Kap86} D. B. Kaplan and A. E. Nelson, {\sl Phys. Lett. {\bf
175B}}, 57 (1986).

\bibitem{Ne87} A. E. Nelson and D. B. Kaplan, {\sl Phys. Lett. {\bf
B192}}, 193 (1987).

\bibitem{Po91} H. D. Politzer and M. B. Wise, {\sl Phys. Lett. {\bf
B273}}, 156 (1991).

\bibitem{Scha91} J. Schaffner, I. N. Mishustin, L. M. Satarov, H.
St\"ocker, and W. Greiner, {\sl Z. Phys. {\bf A341}}, 47 (1991).

\bibitem{Shor85} A. Shor, {\sl Phys. Rev. Lett. {\bf 54}}, 1122 (1985).

\bibitem{Ba91} C. Baglin et al. [NA38 collaboration], {\sl Phys. Lett.
{\bf B272}}, 449 (1991).

\bibitem{Ko90} P. Koch, U. Heinz, and J. Pi\v s\'ut, {\sl Phys. Lett. {\bf
B243}}, 149 (1990); {\sl Z. Phys. {\bf C47}}, 477 (1990).

\bibitem{Gr91} F. Grassi and H. Heiselberg, {\sl Phys. Lett. {\bf
B267}}, 1 (1991).

\bibitem{Koch91} P. Koch, {\sl Prog. Part. Nucl. Phys. {\bf 26}}, 253
(1991).

\bibitem{Centauro} C. M. G. Lattes, Y. Fujimoto, and S. Hasegawa, {\sl
Phys. Rep. {\bf 65}}, 151 (1980); G. Arnison et al., {\sl Phys. Lett.
{\bf 122B}}, 189 (1983); G. J. Alner et al., {\sl Phys. Lett. {\bf
180B}}, 415 (1986), {\sl Phys. Rep. {\bf 154}}, 247 (1987); J. R. Ren
et al., {\sl Phys. Rev. {\bf D38}}, 1417 (1988); L. T. Baradzei et
al., {\sl Nucl. Phys. {\bf B370}}, 365 (1992).

\bibitem{Lam84} C. S. Lam and S. Y. Lo, {\sl Phys. Rev. Lett. {\bf
52}}, 1184 (1984); {\sl Phys. Rev. {\bf D33}}, 1336 (1986).

\bibitem{Pr93} S. Pratt, {\sl Phys. Lett. {\bf B301}}, 159 (1993); S.
Pratt and V. Zelevinsky, {\sl Phys. Rev. Lett. {\bf 72}}, 816 (1994).

\bibitem{Ans89} A. A. Anselm and M. G. Ryskin, {\sl Phys. Lett. {\bf
B266}}, 482 (1989).

\bibitem{Bl92} J.-P. Blaizot and A. Krzywicki, {\sl Phys. Rev. {\bf
D46}}, 246 (1992).

\bibitem{Bj92} J. D. Bjorken, {\sl Acta Phys. Polon. {\bf B23}}, 637
(1992).

\bibitem{Kow92} K. L. Kowalski and C. C. Taylor, preprint CWRUTH-92-6
(1992), $\langle$hep-ph/9211282$\rangle$; J. D. Bjorken, K. L. Kowalski
and C. C. Taylor, preprint SLAC-PUB-6109 (1993).

\bibitem{Ra93} K. Rajagopal and F. Wilczek, {\sl Nucl. Phys. {\bf
B379}}, 395 (1993);  {\sl Nucl. Phys. {\bf B404}}, 577 (1993).

\bibitem{Ga93a} S. Gavin, A. Gocksch and R. D. Pisarski, {\sl Phys.
Rev. Lett. {\bf 72}}, 2143 (1994).

\bibitem{Bo93} D. Boyanovsky, D.-S. Lee, and A. Singh, {\sl Phys.
Rev. {\bf D48}}, 800 (1993); D. Boyanovsky, H. J. deVega, and R.
Holman, preprint PITT-94-01, Univ. of Pittsburgh (1994),
$\langle$hep-ph/9401308$\rangle$.

\bibitem{Be93} P. Bedaque and A. Das, {\sl Mod. Phys. Lett. {\bf A8}},
3151 (1993).

\bibitem{Ga93b} S. Gavin and B. M\"uller, {\sl Phys. Lett. {\bf
B329}}, 486 (1994).

\bibitem{Kl93} S. Yu. Khlebnikov, {\sl Mod. Phys. Lett. {\bf A8}},
1901 (1993).

\bibitem{Ko93} I. I. Kogan, {\sl Phys. Rev. {\bf D48}}, 3971 (1993).

\bibitem{Hu93} Z. Huang and X. N. Wang, {\sl Phys. Rev. {\bf D49}},
4335 (1994).

\bibitem{Bl94} J. P. Blaizot and A. Krzywicki, {\sl Phys. Rev. {\bf
D50}}, 442 (1994).

\bibitem{Klu94} Y. Kluger, preprint LA-UR-94-2754, Los Alamos (1994),
$\langle$hep-ph/9408286$\rangle$.

\bibitem{Asa94} Y. Asakawa, Z. Huang, and X. N. Wang, preprint
LBL-35981, Berkeley (1994), $\langle$hep-ph/9410299$\rangle$.

\bibitem{Gr93} C. Greiner, C. Gong, and B. M\"uller, {\sl Phys. Lett.
{\bf B316}}, 226 (1993).

\bibitem{Co94} T. D. Cohen, M. K. Banerjee, M. Nielsen, and X. M. Jin,
{\sl Phys. Lett. {\bf B333}}, 166 (1994).

\bibitem{Mehr88} F. Karsch, M. T. Mehr, and H. Satz, {\sl Z. Phys.
{\bf C37}}, 617 (1988).

\bibitem{De86} T. A. DeGrand and C. E. DeTar, {\sl Phys. Rev. {\bf
D34}}, 2469 (1986).

\bibitem{Ka86} K. Kanaya and H. Satz, {\sl Phys. Rev. {\bf D34}}, 3193
(1986).

\bibitem{Ka88} F. Karsch and H. W. Wyld, {\sl Phys. Lett. {\bf 213B}},
505 (1988).

\bibitem{KaSa91} F. Karsch and H. Satz, {\sl Z. Phys. {\bf C51}}, 209
(1991).

\bibitem{Cer90} V. \v Cern\'y, I. Horv\'ath, R. Lietava, A. Nogova,
and J. Pi\v sut, {\sl Z. Phys. {\bf C46}}, 481 (1990).

\bibitem{Huf90} J. H\"ufner, B. Povh, and S. Gardner, {\sl Phys. Lett.
{\bf B238}}, 103 (1990).

\bibitem{Thews90} J. Cleymans and R. L. Thews, {\sl Z. Phys. {\bf
C45}}, 391 (1990); R. L. Thews, {\sl Nucl. Phys. {\bf A525}}, 685c
(1991).

\bibitem{Bl91} D. Blaschke, {\sl Nucl. Phys. {\bf A525}}, 269c (1991).

\bibitem{Bl87a} J. P. Blaizot and J. Y. Ollitraut, {\sl Phys. Lett.
{\bf 199B}}, 499 (1987).

\bibitem{Ka88a} F. Karsch and R. Petronzio, {\sl Phys. Lett. {\bf
212B}}, 255 (1988).

\bibitem{Ruu88} P. V. Ruuskanen and H. Satz, {\sl Z. Phys. {\bf C37}},
623 (1988).

\bibitem{Chu88} M. C. Chu and T. Matsui, {\sl Phys. Rev. {\bf D37}},
1851 (1988).

\bibitem{Sat88} H. Satz, {\sl Nucl. Phys. {\bf A488}}, 511 (1988).

\bibitem{Mat88} T. Matsui, {\sl Z. Phys. {\bf C38}}, 245 (1988).

\bibitem{Fta89} J. Ftacnik, P. Lichard, N. Pi\v sutova, and J. Pi\v
s\'ut, {\sl Z. Phys. {\bf C42}}, 139 (1989).

\bibitem{Rop88} G. R\"opke, D. Blaschke, and H. Schultz, {\sl Phys.
Rev. {\bf D38}}, 3589 (1988).

\bibitem{Raha89} S. Raha and B. Sinha, {\sl Phys. Lett. {\bf B218}},
413 (1989).

\bibitem{Lie91} R. Lietava, {\sl Z. Phys. {\bf C50}}, 107 (1991).

\bibitem{Gaz91} M. Ga\'zdzicki and S. Mr\'owczy\'nski, {\sl Z. Phys.
{\bf C49}}, 546 (1991).

\bibitem{Hi90} S. Hioki, T. Kanki and O. Miyamura, {\sl Prog. Theor.
Phys. {\bf 84}}, 317 (1990);  {\bf 85}, 603 (1991).

\bibitem{Ga88} S. Gavin, M. Gyulassy, and A. Jackson, {\sl Phys. Lett.
{\bf 207B}}, 257 (1988).

\bibitem{Vo88} R. Vogt, M. Prakash, P. Koch, and T. H. Hansson, {\sl
Phys. Lett. {\bf 208B}}, 263 (1988).

\bibitem{Ga88b} S. Gavin and M. Gyulassy, {\sl Phys. Lett. {\bf
B214}}, 241 (1988).

\bibitem{Ga90} S. Gavin, R. Vogt, {\sl Nucl. Phys. {\bf B345}}, 104
(1990).

\bibitem{Vo91} R. Vogt, S. J. Brodsky, and P. Hoyer, {\sl Nucl. Phys.
{\bf B360}}, 67 (1991).

\bibitem{Alde91} D. M. Alde et al., {\sl Phys. Rev. Lett. {\bf 66}},
133 (1991).

\bibitem{Ama91} P. Amandruz et al., preprint CERN-PPE-91-198 (1991).

\bibitem{Gup92} S. Gupta and H. Satz, {\sl Phys. Lett. {\bf B283}},
439 (1992).

\bibitem{Vo92} R. Vogt, S. J. Brodsky, and P. Hoyer, {\sl Nucl. Phys.
{\bf B383}}, 643 (1992).

\bibitem{Don93} M. A. Doncheski, M. B. Gay Ducati, and F. Halzen,
preprint MAD-PH-742, Univ. Wisconsin-Madison (1993),
$\langle$hep-ph/9302262$\rangle$.

\bibitem{NA38} C. Baglin et al. [NA38 collaboration], {\sl Phys. Lett.
{\bf B220}}, 471 (1989); {\bf B251}, 465 and 471 (1990); {\bf B262},
362 (1991); {\bf B268}, 453 (1991); {\bf B270}, 105 (1991); M. C.
Abreu et al. [NA38 collaboration], {\sl Nucl. Phys. {\bf A544}}, 209c
(1992).

\bibitem{Gavin93} S. Gavin, H. Satz, R. L. Thews, and R. Vogt,
{\sl Z. Phys. {\bf C61}}, 351 (1994); S. Gavin, preprint BNL 49421,
Brookhaven (1993).

\bibitem{Shen93} J. G. Shen and X. J. Ziu, {\sl Nucl. Phys. {\bf
A55a}}, 708 (1993); {\sl Z. Phys. {\bf C50}}, 85 (1991).

\bibitem{Bj82} J. S. Bjorken, Fermilab publication 82/59, Batavia
(1982), unpublished.

\bibitem{Sve88} B. Svetitsky, {\sl Phys. Rev. {\bf D37}}, 2484 (1988).

\bibitem{Tho91} M. H. Thoma and M. Gyulassy, {\sl Nucl. Phys. {\bf
B351}}, 491 (1991); E. Braaten and M. H. Thoma, {\sl Phys. Rev. {\bf
D44}}, R2625 (1991).

\bibitem{Mro91} S. Mr\'owczy\'nski, {\sl Phys. Lett. {\bf B269}},
383 (1991).

\bibitem{Koi91} Y. Koike and T. Matsui, {\sl Phys. Rev. {\bf D45}},
3237 (1992).

\bibitem{Gyu91} M. Gyulassy, M. Pl\"umer, M. H. Thoma, and X. N. Wang,
{\sl Nucl. Phys. {\bf A538}}, 37c (1992).

\bibitem{Shu78} E. V. Shuryak, {\sl Phys. Lett. {\bf 78B}}, 150
(1978).

\bibitem{MT85} L. D. McLerran and T. Toimela, {\sl Phys. Rev. {\bf
D31}}, 545 (1985).

\bibitem{Raha91} S. Raha and B. Sinha, {\sl Int. J. Mod. Phys. {\bf
A6}}, 517 (1991).

\bibitem{Wel91} H. A. Weldon, {\sl Phys. Rev. Lett. {\bf 66}}, 283
(1991).

\bibitem{Gale91} C. Gale and J. Kapusta, {\sl Phys. Rev. {\bf D43}},
3080 (1991).

\bibitem{Braa90} E. Braaten, R. D. Pisarski, and T. C. Yuan, {\sl
Phys. Rev. Lett. {\bf 64}}, 2242 (1990).

\bibitem{Asa91} M. Asakawa and T. Matsui, {\sl Phys. Rev. {\bf D43}}, 2871
(1991).

\bibitem{Fe76} E. L. Feinberg, {\sl Nuovo Cim. {\bf 34A}}, 391 (1976).

\bibitem{Dom81} G. Domokos and J. I. Goldman, {\sl Phys. Rev. {\bf
D23}}, 203 (1981).

\bibitem{Kaj81} K. Kajantie and H. I. Miettinen, {\sl Z. Phys. {\bf
C9}}, 341 (1981); {\bf C14}, 357 (1982).

\bibitem{Chin82} S. A. Chin, {\sl Phys. Lett. {\bf 119B}}, 51 (1982).

\bibitem{Dom83} G. Domokos, {\sl Phys. Rev. {\bf D28}}, 123 (1983).

\bibitem{McL85} L. McLerran and T. Toimela, {\sl Phys. Rev. {\bf D31}},
545 (1985).

\bibitem{Cley86} J. Cleymans and J. Fingberg, {\sl Phys. Lett. {\bf
168B}}, 405 (1986); J. Cleymans, J. Fingberg and K. Redlich, {\sl
Phys. Rev. {\bf D35}}, 2153 (1987).

\bibitem{Hwa85} R. Hwa and K. Kajantie, {\sl Phys. {\bf D32}}, 1109
(1985).

\bibitem{Cley91} J. Cleymans, K. Redlich, and H. Satz, {\sl Z. Phys.
{\bf C52}}, 517 (1991).

\bibitem{Kaj86} K. Kajantie, M. Kataja, L. McLerran, and P. V.
Ruuskanen, {\sl Phys. Rev. {\bf D34}}, 811 (1986); K. Kajantie, J.
Kapusta, L. McLerran, and A. Mekjian, {\sl Phys. Rev. {\bf D34}}, 2746
(1986).

\bibitem{Kaj87} K. Kajantie, {\sl Nucl. Phys. {\bf A461}}, 225c (1987).

\bibitem{Ruu91} P. V. Ruuskanen, {\sl Nucl. Phys. {\bf A525}}, 255c
(1991).

\bibitem{Ruu92} P. V. Ruuskanen, {\sl Nucl. Phys. {\bf A544}}, 169c
(1992).

\bibitem{Kap92} J. Kapusta, L. McLerran, and D. K. Srivastava, {\sl
Phys. Lett. {\bf B283}}, 145 (1992).

\bibitem{Gei93} K. Geiger and J. I. Kapusta, {\sl Phys. Rev. Lett.
{\bf 70}}, 1920 (1993).

\bibitem{Shu93} E. L. Shuryak and L. Xiong, {\sl Phys. Rev. {\bf 70}},
2241 (1993).

\bibitem{Kam92} B. K\"ampfer and O. P. Pavlenko, {\sl Phys. Lett. {\bf
B289}}, 127 (1992); {\sl Nucl. Phys. {\bf A566}}, 351c (1994).

\bibitem{Kaw92} I. Kawrakow and J. Ranft, preprint UL-HEP-92-08, Univ.
Leipzig (1992).

\bibitem{Stri94} M. T. Strickland, {\sl Phys. Lett. {\bf B331}}, 245
(1994).

\bibitem{Vogt93} R. Vogt, B. V. Jacak, P. L. McGaughey, and P. V.
Ruuskanen, {\sl Phys. Rev. {\bf D49}}, 3345 (1994).

\bibitem{Sar94} I. Sarcevic and P. Valerio, preprint AZPH-TH-94-13,
Univ. of Arizona (1994) $\langle$nucl-th/9405001$\rangle$.

\bibitem{Sie85} P. J. Siemens and S. A. Chin, {\sl Phys. Rev. Lett.
{\bf 55}}, 1266 (1985).

\bibitem{Sei92} D. Seibert, {\sl Phys. Rev. Lett. {\bf 68}}, 1476
(1992).

\bibitem{Kat92} M. Kataja, P. V. Ruuskanen, J. Letessier, and A.
Tounsi, {\sl Z. Phys. {\bf C55}}, 153 (1992).

\bibitem{Hei91} U. Heinz and K. S. Lee, {\sl Phys. Lett. {\bf 259B}},
162 (1991).

\bibitem{Barz91} H. W. Barz, G. Bertsch, B. L. Friman, H. Schulz, and
S. Boggs, {\sl Phys. Lett. {\bf 265B}}, 219 (1991).

\bibitem{Chan92} Z. Aouissat, G. Chanfray, P. Schuck, and G. Welke,
{\sl Z. Phys. {\bf A340}}, 347 (1991); G. Chanfray and P. Schuck, {\sl
Nucl. Phys. {\bf A545}}, 271c (1992).

\bibitem{Ko92} C. M. Ko, P. L\'evai, and W. J. Qiu, {\sl Phys. Rev.
{\bf C46}}, 1159 (1992).

\bibitem{Herr92} M. Herrmann, B. L. Friman, and W. N\"orenberg, {\sl
Z. Phys. {\bf A343}}, 119 (1992).

\bibitem{Hat93} T. Hatsuda and Y. Koike, {\sl Nucl. Phys. {\bf B394}},
221 (1993).

\bibitem{Hag93} K. L. Haglin and C. Gale, {\sl Nucl. Phys. {\bf
B421}}, 613 (1994).

\bibitem{Lis91} D. Lissauer and E. V. Shuryak, {\sl Phys. Lett. {\bf
B253}}, 15 (1991).

\bibitem{Bi91} P. Z. Bi and J. Rafelski, {\sl Phys. Lett. {\bf B262}},
485 (1991).

\bibitem{Asa93} M. Asakawa and C. M. Ko, {\sl Phys. Lett. {\bf B322}},
33 (1994).

\bibitem{WA80} R. Albrecht et al. [WA80 collaboration], {\sl Z.
Phys. {\bf C51}}, 1 (1991); {\sl Nucl. Phys. {\bf A566}}, 61c (1993).

\bibitem{Sri92} D. K. Srivastava, B. Sinha, M. Gyulassy, and X. N.
Wang, {\sl Phys. Lett. {\bf B276}}, 285 (1992).

\bibitem{KLS} J. Kapusta, P. Lichard, and D. Seibert, {\sl Phys. Rev.
{\bf D44}}, 2774 (1991); {\sl Nucl. Phys. {\bf A544}}, 485c (1992);
{\sl Phys. Rev. {\bf D47}}, 4171(E) (1993).

\bibitem{Xiong92} L. Xiong, E. Shuryak, and G. E. Brown, {\sl Phys.
Rev. {\bf D46}}, 3798 (1992).

\bibitem{Bai92} R. Baier, H. Nakkagawa, A. Niegawa, and K. Redlich,
{\sl Z. Phys. {\bf C53}}, 433 (1992); {\sl Phys. Rev. {\bf D45}}, 4323
(1992).

\bibitem{Cha92} S. Chakrabarty, J. Alam, D. K. Srivastava, B. Sinha,
and S. Raha, {\sl Phys. Rev. {\bf D46}}, 3802 (1992).

\bibitem{Sei93} D. Seibert, {\sl Z. Phys. {\bf C58}}, 307 (1993).

\bibitem{Alam93} J. Alam, D. K. Srivastava, B. Sinha, and D. N. Basu,
{\sl Phys. Rev. {\bf D48}}, 1117 (1993).

\bibitem{Neu94} J. J. Neumann, D. Seibert, and G. Fai, preprint
KSUNCR-016-94, Kent State University (1994),
$\langle$nucl-th/9409008$\rangle$.


\bibitem{Yo87} R. Yoshida, T. Miyazaki, and M. Kadoya, {\sl Phys. Rev.
{\bf D35}}, 388 (1987).

\bibitem{Red87} K. Redlich, {\sl Phys. Rev. {\bf D36}}, 3378 (1987).

\bibitem{Mak89} A. N. Makhlin, {\sl Yad. Fiz. {\bf 49}}, 238 (1989).

\bibitem{Sri93a} D. K. Srivastava and J. I. Kapusta, {\sl Phys. Lett.
{\bf B307}}, 1 (1993).

\bibitem{Sri93b} D. K. Srivastava and J. I. Kapusta, {\sl Phys. Rev.
{\bf C48}}, 1335 (1993); D. K. Srivastava and C. Gale, {\sl Phys.
Lett. {\bf 319}}, 407 (1993).

\bibitem{Bod71} A. Bodmer, {\sl Phys. Rev. {\bf D4}}, 1601 (1971).

\bibitem{FM78} B. Friedman and L. McLerran, {\sl Phys. Rev. {\bf
D17}}, 1109 (1978).

\bibitem{Chin79} S. A. Chin and A. K. Kerman, {\sl Phys. Rev. Lett.
{\bf 43}}, 1292 (1979).

\bibitem{Wit84} E. Witten, {\sl Phys. Rev. {\bf D30}}, 272 (1984).

\bibitem{Far84} E. Farhi and R. L. Jaffe, {\sl Phys. Rev. {\bf D30}},
2379 (1984).

\bibitem{Kas93} M. Kasuya, T. Saito, and M. Yasue, {\sl Phys. Rev.
{\bf D47}}, 2153 (1993).

\bibitem{Gil93} E. P. Gilson and R. L. Jaffe, {\sl Phys. Rev. Lett.
{\bf 71}}, 332 (1993).

\bibitem{Mad93} J. Madsen, {\sl Phys. Rev. Lett. {\bf 70}}, 391
(1993); {\sl Phys. Rev. {\bf D47}}, 5156 (1993).

\bibitem{Des93} M. S. Desai, H. J. Crawford, and G. L. Shaw, {\sl
Phys. Rev. {\bf D47}}, 2063 (1993).

\bibitem{Cha93a} S. Chakrabarty, {\sl Phys. Rev. {\bf D48}}, 1409
(1993).

\bibitem{Lee93} K. S. Lee and U. Heinz, {\sl Phys. Rev. {\bf D47}},
2068 (1993).

\bibitem{Mad88} J. Madsen, {\sl Phys. Rev. Lett. {\bf 61}}, 2909
(1988).

\bibitem{Hei92} H. Heiselberg, {\sl Phys. Scr. {\bf 46}}, 485 (1992).

\bibitem{Craw94} H. J. Crawford and C. H. Greiner, {\sl Scientific
American {\bf 270}}, 58 (1994).

\bibitem{Liu84} H. Liu and G. L. Shaw, {\sl Phys. Rev. {\bf D30}},
1137 (1984).

\bibitem{Grei87} C. Greiner, P. Koch, and H. St\"ocker, {\sl Phys.
Rev. Lett. {\bf 58}}, 1109 (1978); C. Greiner, D. Rischke, H.
St\"ocker, and P. Koch, {\sl Phys. Rev. {\bf D38}}, 2797 (1988); C.
Greiner and H. St\"ocker, {\sl Phys. Rev. {\bf D44}}, 3517 (1991).

\bibitem{Shaw92} H. J. Crawford, M. S. Desai, and G. L. Shaw, {\sl
Phys. Rev. {\bf D45}}, 857 (1992).

\bibitem{Bar90} J. Barrette et al. [E-814 collaboration], {\sl Phys.
Lett. {\bf B252}}, 550 (1990).

\bibitem{Aoki92} M. Aoki et al. [E-858 collaboration], {\sl Phys. Rev.
Lett. {\bf 69}}, 2345 (1992).

\bibitem{Bor94} K. Borer et al. [{\sc Newmass} collaboration], {\sl Phys.
Rev. Lett. {\bf 72}}, 1415 (1994).

\bibitem{Long94} R. S. Longacre et al. [E-810 collaboration], {\sl
Nucl. Phys. {\bf A566}}, 167c (1994).

\bibitem{App94} G. Appelquist et al. [NA52 collaboration], {\sl Nucl.
Phys. {\bf A566}}, 507c (1994).

\bibitem{Grei89} C. Greiner and B. M\"uller, {\sl Phys. Lett. {\bf
B219}}, 199 (1989).

\bibitem{Gyu92a} M. Gyulassy, {\sl Phys. Lett. {\bf B286}}, 211 (1992).

\bibitem{Scha93} J. Schaffner, C. B. Dover, A. Gal, C. Greiner, and H.
St\"ocker, {\sl Phys. Rev. Lett. {\bf 71}}, 1328 (1993).

\bibitem{Wer94} K. Werner, {\sl Phys. Rev. Lett. {\bf 73}}, 1594
(1994).

\end{references}
\end{document}